\documentclass[11pt]{article}

\usepackage{amsmath}
\usepackage{bm}
\usepackage{amssymb}
\usepackage{amstext}
\usepackage{latexsym}
\usepackage{color}
\usepackage{graphicx}
\usepackage{epsfig}
\usepackage{subfigure}
\usepackage{rotating}
\usepackage{curves}
\usepackage{epic}
\usepackage{amsthm}

\usepackage{amsfonts}
\usepackage[T1]{fontenc}
\usepackage[latin1]{inputenc}
\usepackage{authblk}
\usepackage{tikz}
\usetikzlibrary{shapes,arrows}
\usepackage{graphics}
\usepackage{verbatim}
\usepackage{color}
\usepackage{hyperref}

\newcommand{\A}{{\mathfrak A}}

\newcommand{\mc}{\mathcal}

\newcommand{\be}{\begin{equation}}
\newcommand{\en}{\end{equation}}
\newcommand{\D}{{\mc D}}

\newcommand{\1}{1 \!\! 1}

\newtheorem{thm}{Theorem}
\newtheorem{cor}[thm]{Corollary}

\newtheorem{defi}{Definition}[section]
\newtheorem{lem}[defi]{Lemma}

\newtheorem{Theo}{Theorem}[section]

\newtheorem{remark}[Theo]{Remark}

\newcommand{\bedefin}{\begin{defi}}
\newcommand{\findefi}{\end{defi} \medskip}

\newcommand{\betheo}{\begin{theorem}$\!\!${\bf \,\,\,}}
\newcommand{\entheo}{\end{theorem}}
\newcommand{\enth}{\end{theorem}}

\newcommand{\becor}{\begin{cor}$\!\!${\bf .}}
\newcommand{\encor}{\end{cor}}

\newcommand{\belem}{\begin{lem}$\!\!${\bf }}
\newcommand{\enlem}{\end{lem}}

\newcommand{\prf}{\noindent{\bf{ Proof}\,\,}}

\newcommand{\bea}{\begin{eqnarray}}
\newcommand{\ena}{\end{eqnarray}}

\newcommand{\beano}{\begin{eqnarray*}}
\newcommand{\enano}{\end{eqnarray*}}

\newcommand{\bee}{\begin{enumerate}}
\newcommand{\ene}{\end{enumerate}}

\newcommand{\bei}{\begin{itemize}}
\newcommand{\eni}{\end{itemize}}

\newcommand{\betab}{\begin{tabular}}
\newcommand{\entab}{\end{tabular}}

\newcommand{\bd}{\begin{displaymath}}

\newcommand{\C}{{\mathbb C}}

\newcommand{\bp}{\mathbf p}
\newcommand{\bq}{\mathbf q}

\newcommand{\g}{G_{\hbox{\tiny{NC}}}}
\newcommand{\G}{\mathfrak{g}_{\hbox{\tiny{NC}}}}

\catcode `\@=11 \@addtoreset{equation}{section}

\catcode `\@=12

\begin{document}

\title{Triply Extended Group of Translations of $\mathbb{R}^{4}$ as Defining Group of NCQM: relation to various gauges}
\author{S. Hasibul Hassan Chowdhury$^\dag$ \\
  and S. Twareque Ali$^{\dag\dag}$\\
\medskip
\small{Department of Mathematics and Statistics,\\ Concordia University, Montr\'eal, Qu\'ebec, Canada H3G 1M8}\\

{\footnotesize $^\dag$e-mail: schowdhury@mathstat.concordia.ca\\
 $^{\dag\dag}$e-mail: twareque.ali@concordia.ca}}

\date{\today}

\maketitle

\begin{abstract}
The role of the triply extended group of translations of $\mathbb{R}^{4}$, as the defining group of two dimensional noncommutative quantum mechanics (NCQM), has been studied in \cite{ncqmjmp}. In this paper, we revisit the coadjoint orbit structure and various irreducible representations of the group associated with them. The two irreducible representations corresponding to the Landau and symmetric gauges are found to arise from the underlying defining group. The group structure of the transformations, preserving the commutation relations of NCQM, has been studied along with specific examples. Finally, the relationship of a certain family of  UIRs of the underlying defining group with a family of deformed complex Hermite polynomials has been explored.

\end{abstract}

\section{Introduction}\label{sec:intro}
Noncommutative quantum mechanics is an active area of research these days.
The thinking here is that a  modification  of ordinary quantum mechanics
is needed to model physical space-time at very short distances.
One way to bring about this modification is by an alteration  of the canonical
commutation relations  of standard quantum mechanics. A commonly studied version of such a modification looks at a quantum system with two degrees of freedom. The standard quantum mechanics of such a system is governed by the commutation relations
\be
  [Q_i , P_j ] = i\hbar \delta_{ij}I,\; \; i, j = 1,2,
\label{nc-comm-relns1}
\en
Here the $Q_i, P_j$ are the quantum mechanical position and momentum observables, respectively.
In non-commutative quantum mechanics one starts by imposing the additional commutation relation
\be
[Q_1, Q_2] = i\vartheta I,
\label{nc-comm-relns10}
\en
where $\vartheta$ is a small, positive parameter  which measures the additionally introduced noncommutativity between the observables of the two spatial coordinates.
The limit $\vartheta = 0$ then corresponds to standard (two-dimensional) quantum mechanics. One could continue by also making the two momenta noncommuting:
\be
  [P_i , P_j ] = i\gamma\epsilon_{ij} I\; , \qquad i,j = 1,2\; ,\quad \epsilon_{11} =
  \epsilon_{22} = 0, \; \epsilon_{12} = -\epsilon_{21} = 1\; ,
\label{non-comm-relns2}
\en
where $\gamma$ is yet another positive parameter. In standard quantum mechanics, the presence of such a non-zero commutator would mean that there is a magnetic field in the system. Such models have been studied extensively in the recent literature, from different perspectives, e.g., its relationship to noncommutative geometry, deformation quantization, the Moyal product, the question of spatial extension and localization, the Landau problem of an electron in a constant magnetic field, etc. As a small sampling of that literature, we may mention  \cite{acat,bas,ncqmjmp,delducetalt,dias,gamboa1,hormarsti,scho2,scho}, a  list that is by no means complete; however, the interested reader will find a wealth of additional references in it.

Our aim in this paper is to  study in some detail the underlying group theoretical structure of such a noncommutative system, with two degrees of freedom, and is a sequel to an earlier paper \cite{ncqmjmp} in the same spirit.

The Weyl-Heisenberg group, whose generators in a unitary irreducible representation on a Hilbert space give the position and momentum operators of standard quantum mechanics, satisfying the canonical commutation relations, can be thought of as being the defining group of standard non-relativistic quantum mechanics.
The analog of this group, in the setting of a two-dimensional noncommutative quantum system (i.e., a system in which the two operators of position  are also non-commuting), was explored in \cite{ncqmjmp}. There the possibility of an additional non-commutativity (that of the two momentum operators as well) was also  considered.
In the literature (see, for example, \cite{delducetalt}), two different gauges and their physical interpretations have been pointed out, connected with this latter non-commutativity (of the momenta). We shall show in this paper that the irreducible representations of the resulting commutation relations, postulated there arise indeed from the irreducible unitary representations of the triply extended group of translations of $\mathbb{R}^{4}$, which we shall denote as $\g$ from now on. (In \cite{ncqmjmp} the somewhat cumbersome notation $\overline{\overline{\overline{G}}}_T$ had been used for this group). In this sense it is this group which is the defining group of noncommutative quantum mechanics. Indeed, as will be shown in the sequel, the different unitary irreducible representations of it describe all the different levels of  non-commutativity  presently considered in the literature.

In this paper, we give a complete description of all the unitary irreducible representations of the group $\g$ and its Lie algebra $\G$, following the classification of the underlying coadjoint orbits. The unitary irreducible representations of the $2$ dimensional Weyl-Heisenberg group are found to be sitting inside the unitary dual of $\g$. We compute the unitary irreducible representations, associated with the {\em Landau gauge} and the {\em symmetric gauge} of $\g$, explicitly. The transformation group, preserving the commutation relations of noncommutative quantum mechanics, is studied along with an example related to the matrix of transformation between the UIRs of $\G$ in the Landau and symmetric gauges. Finally, we obtain a family of coadjoint orbits in $\G^{*}$ that gives rise to the representations associated to the deformed complex Hermite polynomials studied at length in (\cite{alibaloghshah}, \cite{aliismailshah}).

\section{Coadjoint orbits and UIRs of \texorpdfstring{$\g$}{GNC}}\label{sec:coad-orbts}

The phase space of a free classical system, moving in two spatial dimensions, is the four dimensional abelian group of translations of $\mathbb R^4$. Let us denote a general element of this group by $(\bq,\bp)$, in terms of the two-vectors of position and momentum, respectively. A generic element of the triple central extension $\g$ of this abelian group will be  denoted by $(\theta, \phi, \psi, \bq, \bp)$. The group composition law for $\g$ reads (see \cite{ncqmjmp})
\begin{eqnarray}\label{grp-law}
\lefteqn{(\theta,\phi,\psi,\bq,\bp)(\theta^{\prime},\phi^{\prime},\psi^{\prime},\bq^{\prime},\bp^{\prime})}\nonumber\\
&&=(\theta+\theta^{\prime}+\frac{\alpha}{2}[\langle\bq,\bp^{\prime}\rangle-\langle\bp,\bq^{\prime}\rangle],\phi+\phi^{\prime}+\frac{\beta}{2}[\bp\wedge\bp^{\prime}],\psi+\psi^{\prime}+\frac{\gamma}{2}[\bq\wedge\bq^{\prime}]\nonumber\\
&&\;\;\;,\bq+\bq^{\prime},\bp+\bp^{\prime}),
\end{eqnarray}
where $\alpha$, $\beta$ and $\gamma$ are certain strictly positive dimensionful constants associated with the central extensions corresponding to $\theta$, $\phi$, and $\psi$, respectively. Also, $\bq = (q_1, q_2)$  and $\bp = (p_{1},p_{2})$. In (\ref{grp-law}), $\langle.,.\rangle$ and $\wedge$ are defined as $\langle {\bq},{\bp}\rangle:=q_{1}p_{1}+q_{2}p_{2}$ and ${\bq}\wedge{\bp}:=q_{1}p_{2}-q_{2}p_{1}$ respectively.

If we denote the dimension of the momentum coordinate by $[p]$ and that of the position coordinate by $[q]$, then we immediately see from (\ref{grp-law}) that in order to have $\theta$, $\phi$ and $\psi$ to be all dimensionless, we must have
\begin{equation}\label{dimensional-analysis}
[\alpha]=\left[\frac{1}{pq}\right],\;\;[\beta]=\left[\frac{1}{p^2}\right],\;\;\hbox{and}\;\;[\gamma]=\left[\frac{1}{q^2}\right].
\end{equation}

A matrix realization of $\g$ is as follows
\begin{equation}\label{mtrx-relztn}
(\theta,\phi,\psi,\bq,\bp)_{\alpha,\beta,\gamma}=\begin{bmatrix}1&0&0&-\frac{\alpha}{2}p_{1}&-\frac{\alpha}{2}p_{2}&\frac{\alpha}{2}q_{1}&\frac{\alpha}{2}q_{2}&\theta\\0&1&0&0&0&-\frac{\beta}{2}p_{2}&\frac{\beta}{2}p_{1}&\phi\\0&0&1&-\frac{\gamma}{2}q_{2}&\frac{\gamma}{2}q_{1}&0&0&\psi\\0&0&0&1&0&0&0&q_{1}\\0&0&0&0&1&0&0&q_{2}\\0&0&0&0&0&1&0&p_{1}\\0&0&0&0&0&0&1&p_{2}\\0&0&0&0&0&0&0&1\end{bmatrix}.
\end{equation}

Let us denote the Lie algebra of $\g$ by $\G$. If we denote the basis elements of $\G$ by $\Theta,\Phi,\Psi,Q_{1},Q_{2},P_{1}$ and $P_{2}$, corresponding to the on-parameter subgroups generated by the group parameters $\theta,\phi,\psi,p_{1},p_{2},q_{1}$ and $q_{2}$, respectively, we end up with the following Lie bracket relations between them
\begin{equation}\label{commut-reltns}
\begin{split}
&[P_{i},Q_{j}]=\alpha\delta_{i,j}\Theta,\quad
 [Q_{1},Q_{2}]=\beta\Phi,\quad
 [P_{1},P_{2}]=\gamma\Psi,\quad
 [P_{i},\Theta]=0,\\
&[Q_{i},\Theta]=0,\quad
 [P_{i},\Phi]=0,\quad
 [Q_{i},\Phi]=0,\quad
 [P_{i},\Psi]=0,\\
&[Q_{i},\Psi]=0,\quad
 [\Theta,\Phi]=0,\quad
 [\Phi,\Psi]=0,\quad
 [\Theta,\Psi]=0, \quad i,j =1,2\; .
\end{split}
\end{equation}

For the sake of later convenience, we switch to a different notation by replacing the group parameters $p_{1}$, $p_{2}$, $q_{1}$, $q_{2}$, $\theta$, $\phi$ and $\psi$ with $a_{1}$, $a_{2}$, $a_{3}$, $a_{4}$, $a_{5}$, $a_{6}$ and $a_{7}$, respectively. Using matrix representations, the respective one parameter group generators will be  denoted by $\mathcal{X}_{1}$, $\mathcal{X}_{2}$, $\mathcal{X}_{3}$, $\mathcal{X}_{4}$, $\mathcal{X}_{5}$, $\mathcal{X}_{6}$ and $\mathcal{X}_{7}$, respectively. The matrix representation $\mathcal{X}$ of a generic Lie algebra element $X\in\G$ then reads
\begin{equation*}
\mathcal{X}=x^{1}\mathcal{X}_{1}+x^{2}\mathcal{X}_{2}+x^{3}\mathcal{X}_{3}+x^{4}\mathcal{X}_{4}+x^{5}\mathcal{X}_{5}+x^{6}\mathcal{X}_{6}+x^{7}\mathcal{X}_{7},
\end{equation*}
with
\begin{equation}\label{algbr-reprstn}
\mathcal{X}=\begin{bmatrix}0&0&0&-\frac{\alpha}{2}x^{1}&-\frac{\alpha}{2}x^{2}&\frac{\alpha}{2}x^{3}&\frac{\alpha}{2}x^{4}&x^{5}\\0&0&0&0&0&-\frac{\beta}{2}x^{2}&\frac{\beta}{2}x^{1}&x^{6}\\0&0&0&-\frac{\gamma}{2}x^{4}&\frac{\gamma}{2}x^{3}&0&0&x^{7}\\0&0&0&0&0&0&0&x^{3}\\0&0&0&0&0&0&0&x^{4}\\0&0&0&0&0&0&0&x^{1}\\0&0&0&0&0&0&0&x^{2}\\0&0&0&0&0&0&0&0\end{bmatrix}.
\end{equation}
Under the above mentioned change in notation, a generic group element of $\g$ is now represented by the following matrix:
\begin{equation}\label{modfd-grp-rep}
g(a_{1},a_{2},a_{3},a_{4},a_{5},a_{6},a_{7})=\begin{bmatrix}1&0&0&-\frac{\alpha}{2}a_{1}&-\frac{\alpha}{2}a_{2}&\frac{\alpha}{2}a_{3}&\frac{\alpha}{2}a_{4}&a_{5}\\0&1&0&0&0&-\frac{\beta}{2}a_{2}&\frac{\beta}{2}a_{1}&a_{6}\\0&0&1&-\frac{\gamma}{2}a_{4}&\frac{\gamma}{2}a_{3}&0&0&a_{7}\\0&0&0&1&0&0&0&a_{3}\\0&0&0&0&1&0&0&a_{4}\\0&0&0&0&0&1&0&a_{1}\\0&0&0&0&0&0&1&a_{2}\\0&0&0&0&0&0&0&1\end{bmatrix}.
\end{equation}

Let us write down a generic Lie algebra element $X$ in terms of the abstract basis elements $X_{1}$, $X_{2}$, $X_{3}$, $X_{4}$, $X_{5}$, $X_{6}$ and $X_{7}$ as $X=x^{1}X_{1}+x^{2}X_{2}+x^{3}X_{3}+x^{4}X_{4}+x^{5}X_{5}+x^{6}X_{6}+x^{7}X_{7}$. Here $X_{j}$'s are treated just as monomials and not  as matrices. Now, if we denote the dual lie algebra by $\G^{*}$, then an element $F$ of it can conveniently be represented by the following lower triangular matrix:
\begin{equation}\label{dual-algbra-rep}
F=\begin{bmatrix}0&0&0&0&0&0&0&0\\0&0&0&0&0&0&0&0\\0&0&0&0&0&0&0&0\\0&0&0&0&0&0&0&0\\0&0&0&0&0&0&0&0\\0&0&0&0&0&0&0&0\\0&0&0&0&0&0&0&0\\X_{5}&X_{6}&X_{7}&X_{3}&X_{4}&X_{1}&X_{2}&0\end{bmatrix},
\end{equation}
with the dual pairing being given by $\langle F,X\rangle=\hbox{tr}(F\mathcal{X})=\displaystyle\sum\limits_{i=1}^{7}x^{i}X_{i}$.
We, therefore, note that
\begin{eqnarray}\label{coad-act-derv}
\lefteqn{g(a_{1},a_{2},a_{3},a_{4},a_{5},a_{6},a_{7})Fg(a_{1},a_{2},a_{3},a_{4},a_{5},a_{6},a_{7})^{-1}}\nonumber\\
&&=\begin{bmatrix}*&*&*&*&*&*&*&*\\ *&*&*&*&*&*&*&*\\ *&*&*&*&*&*&*&*\\ *&*&*&*&*&*&*&*\\ *&*&*&*&*&*&*&*\\ *&*&*&*&*&*&*&*\\ *&*&*&*&*&*&*&*\\X_{5}^{\prime}&X_{6}^{\prime}&X_{7}^{\prime}&X_{3}^{\prime}&X_{4}^{\prime}&X_{1}^{\prime}&X_{2}^{\prime}&*\end{bmatrix},\label{coadactionfordoubleexten}
\end{eqnarray}
with
\begin{equation}\label{coad-action-rel}
\begin{split}
&X_{1}^{\prime}=X_{1}-\frac{\alpha}{2}a_{3}X_{5}+\frac{\beta}{2}a_{2}X_{6},\;\;
 X_{2}^{\prime}=X_{2}-\frac{\alpha}{2}a_{4}X_{5}-\frac{\beta}{2}a_{1}X_{6},\\
&X_{3}^{\prime}=X_{3}+\frac{\gamma}{2}a_{4}X_{7}+\frac{\alpha}{2}a_{1}X_{5},\;\;
 X_{4}^{\prime}=X_{4}-\frac{\gamma}{2}a_{3}X_{7}+\frac{\alpha}{2}a_{2}X_{5},\\
&X_{5}^{\prime}=X_{5},\;\;
 X_{6}^{\prime}=X_{6},\;\;
 X_{7}^{\prime}=X_{7}.
\end{split}
\end{equation}
The entries, denoted by $*$'s in (\ref{coad-act-derv}), are of no interest for the present computations.
Thus we arrive at the required coadjoint action of the group on the dual algebra, given by
\begin{eqnarray}\label{coad-action-exprssn}
\lefteqn{Kg(a_{1},a_{2},a_{3},a_{4},a_{5},a_{6},a_{7})(X_{1},X_{2},X_{3},X_{4},X_{5},X_{6},X_{7})}\nonumber\\
&&=(X_{1}-\frac{\alpha}{2}a_{3}X_{5}+\frac{\beta}{2}a_{2}X_{6},\;\;X_{2}-\frac{\alpha}{2}a_{4}X_{5}-\frac{\beta}{2}a_{1}X_{6},\nonumber\\
&&\;\;\;\;X_{3}+\frac{\gamma}{2}a_{4}X_{7}+\frac{\alpha}{2}a_{1}X_{5},\;X_{4}-\frac{\gamma}{2}a_{3}X_{7}+\frac{\alpha}{2}a_{2}X_{5},\;X_{5},\;X_{6},\;X_{7}).
\end{eqnarray}

Let us pause briefly and study the geometry of the relevant coadjoint orbits before computing all unitary irreducible representations of $\g$. From (\ref{coad-action-exprssn}), we immediately see that $X_{5}$, $X_{6}$ and $X_{7}$ belonging to the center of $\G$ remain invariant under the coadjoint action of $\g$, as expected. These three invariant coordinates on the right side of (\ref{coad-action-exprssn}) refer to the $\g$-invariant polynomial functions on $\G^{*}$ related to $X_{5}, X_{6}, X_{7}$, respectively, all belonging to to the center $\mathcal{Z}(\G)$. Let us denote these three polynomial invariants by $P(F)=X_{5}$, $Q(F)=X_{6}$ and $R(F)=X_{7}$.

Now it is important to note that the following system of linear equations in $a_1$, $a_2$, $a_3$ and $a_4$ is solvable:

\begin{equation}\label{simultaneous-solution}
\begin{aligned}
&X_{1}-\frac{\alpha}{2}a_{3}X_{5}+\frac{\beta}{2}a_{2}X_{6}&=&0,\\
&X_{4}-\frac{\gamma}{2}a_{3}X_{7}+\frac{\alpha}{2}a_{2}X_{5}&=&0,\\
&X_{2}-\frac{\alpha}{2}a_{4}X_{5}-\frac{\beta}{2}a_{1}X_{6}&=&0,\\
&X_{3}+\frac{\gamma}{2}a_{4}X_{7}+\frac{\alpha}{2}a_{1}X_{5}&=&0,
\end{aligned}
\end{equation}
if and only if the determinant of the $2\times 2$ matrix

\begin{equation}\label{important-matrix}
W=\begin{bmatrix}\alpha X_{5}&\beta X_{6}\\\gamma X_{7}&\alpha X_{5}\end{bmatrix}
\end{equation}
 is nonzero, i.e.
\begin{equation}\label{determinantal-eqn}
\det{W}=\alpha^{2}X_{5}^{2}-\gamma\beta X_{6}X_{7}\neq 0.
\end{equation}

Also, if we denote the three polynomial invariants $X_5$, $X_6$ and $X_7$ with $\rho$, $\sigma$ and $\tau$, respectively, then the triple $(\rho, \sigma, \tau)$ solely determines the geometry of the underlying coadjoint orbits. For all three of  $\rho$, $\sigma$ and $\tau$ assuming non-zero values and $\det{W}\neq 0$ , the vector $(0,0,0,0,\rho,\sigma,\tau)$ will  always lie in the underlying coadjoint orbit of codimension 3. In other words, these coadjoint orbits will be  $4$ dimensional and will pass through the point $(0,0,0,0,\rho,\sigma,\tau)$ of the dual algebra space $\G^{*}$. Let us denote these 4-dimensional coadjoint orbits by $\mathcal{O}^{\rho,\sigma,\tau}_{4}$:
\begin{equation}\label{nonzero-det-nonzero}
\mathcal{O}^{\rho,\sigma,\tau}_{4}=\{F\in\G^{*}\mid P(F)=\rho,Q(F)=\sigma,R(F)=\tau\}.
\end{equation}

\begin{figure}[thb]
\centering
\includegraphics[width=8cm]{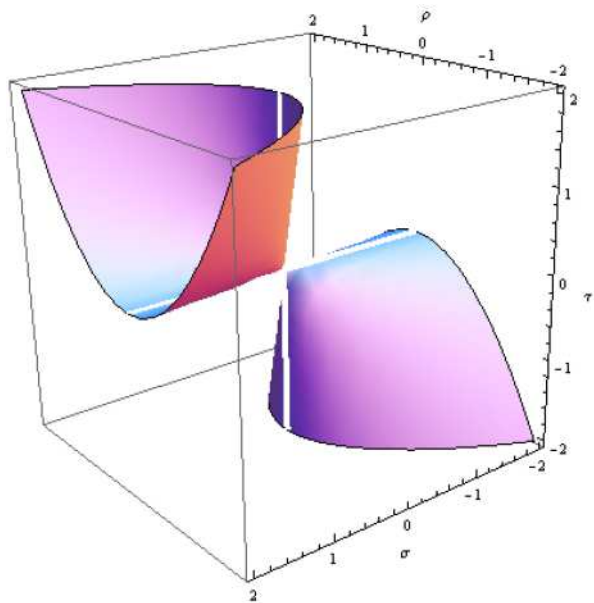}
\caption{The surface $\mathbb{S}_{\rho,\zeta}$ in $\mathbb{R}^{3}_{0}$ that is associated with a family of 2-dimensional coadjoint orbits in the dual Lie algebra. Here, $\alpha$, $\beta$ and $\gamma$ are all taken to be 1 in their appropriate units.}
\label{fig:figfirst}
\end{figure}

A generic element of $\mathcal{O}^{\rho,\sigma,\tau}_{4}$ can be written as $(k_{1}, k_{2}, k_{3}, k_{4}, \rho, \sigma, \tau)$, where $(k_{1}, k_{2}, k_{3},k_{4})$ takes values in $\mathbb{R}^{4}$. These nonintersecting four dimensional coadjoint orbits (one for each choice of the nonzero values of $\rho, \sigma, \tau$ with $\alpha^{2}{\rho}^{2}-\gamma\beta\sigma\tau\neq 0$) are sitting inside the $7$-dimensional dual Lie algebra $\G^{*}$ in the following way. $\mathbb{R}^{7}$ can be regarded as a continuum of nonintersecting $\mathbb{R}^{4}$ spaces going through each point of an $\mathbb{R}^{3}$ space embedded in $\mathbb{R}^{7}$. Let us denote a generic point of the embedded $\mathbb{R}^{3}$ space by $(0, 0, 0, 0, \rho, \sigma, \tau)$. Restricting $\rho$, $\sigma$, and $\tau$ to nonzero real values, we obtain a disconnected toplogical space in the usual Euclidean topology with $8$ connected components which we denote as $\mathbb{R}^{3}_{0}$. Inside $\mathbb{R}^{3}_{0}$, is embedded an elliptic cone-shaped surface (with two perpendicular lines deleted) given by the equation $\alpha^{2}{\rho}^{2}-\gamma\beta\sigma\tau=0$. Any point on such a surface is completely determined by a family of straight lines given by $\rho=\sigma\zeta=\frac{\gamma\beta\tau}{\zeta\alpha^{2}}$ with $\zeta\in(-\infty,0)\cup(0,\infty)$. It can easily be seen that all these straight lines pass through the origin. In figure \ref{fig:figfirst}, this surface, denoted by $\mathbb{S}_{\rho,\zeta}$, is illustrated as being embedded in the submanifold $\mathbb{R}^{3}$ of the dual Lie algebra $\G^{*}$. For any vector $(0,0,0,0,\rho,\sigma,\tau)\in\mathbb{R}^{3}_{0}\setminus \mathbb{S}_{\rho,\zeta}$, an $\mathbb{R}^{4}$ coadjoint orbit $\mathcal{O}^{\rho, \sigma, \tau}_{4}$ passes through a point $(0, 0, 0, 0, \rho, \sigma, \tau)\in\G^{*}$.

Let us, now, consider the case when $(0,0,0,0,\rho,\sigma,\tau)\in\mathbb{S}_{\rho,\zeta}$, with $\sigma=\frac{\rho}{\zeta}$ and $\tau=\frac{\zeta\alpha^{2}\rho}{\gamma\beta}$ for any $\rho\neq 0$ and $\zeta\in(-\infty,0)\cup(0,\infty)$. In this case, the linear system (\ref{simultaneous-solution}) does not admit any solution. Nevertheless, we can always satisfy the first and the third equations of (\ref{simultaneous-solution}) by choosing appropriate $a_1$, $a_2$, $a_3$ and $a_4$. This, then, gives us $2$ new {\em rational invariants} which are as follows

\begin{equation}\label{new-invariants}
\begin{aligned}
&X_{3}+\frac{\alpha X_{5}X_{2}}{\beta X_{6}}&=&\kappa,\\
&X_{4}-\frac{\alpha X_{5}X_{1}}{\beta X_{6}}&=&\delta.
\end{aligned}
\end{equation}

Gathering all $5$ invariants, one immediately deduces from (\ref{coad-action-exprssn}) that the underlying coadjoint orbits pass through the points $(0,0,\kappa,\delta,\rho,\frac{\rho}{\zeta},\frac{\zeta\alpha^{2}\rho}{\gamma\beta})$. These orbits are all of codimension $5$ and hence they are $2$-dimensional. We shall be denoting the underlying coadjoint orbits by $^{\kappa,\delta}\mathcal{O}^{\rho,\zeta}_{2}$, henceforth. A generic point of such an orbit (for fixed $\kappa$, $\delta$, $\rho$ and $\zeta$) is given by $(k_{1},k_{2},\kappa,\delta,\rho,\frac{\rho}{\zeta},\frac{\zeta\alpha^{2}\rho}{\gamma\beta})$ where both $k_{1}$ and $k_{2}$ take their values from the real line. It is important to note that for each $(0,0,0,0,\rho,\frac{\rho}{\zeta},\frac{\zeta\alpha^{2}\rho}{\gamma\beta})\in\mathbb{S}_{\rho,\zeta}$, we can consider an $\mathbb{R}^{4}$ space passing through it inside the $7$-dimensional dual Lie algebra $\G^{*}$. Inside each such $\mathbb{R}^{4}$ space, the $2$-dimensional coadjoint orbits $^{\kappa,\delta}\mathcal{O}^{\rho,\zeta}_{2}$ are foliated with $(\kappa,\delta)\in\mathbb{R}^{2}$.

Now we consider the rest of the points $(0, 0, 0, 0, \rho, \sigma, \tau)$ of the underlying $\mathbb{R}^{3}$ space embedded in $\mathbb{R}^{7}$ and denote this set by $\mathbb{R}^{3}_{1}$. Let us subdivide the points belonging to $\mathbb{R}^{3}_{1}$ into the following classes
\begin{itemize}
\item [$ \bullet $] The points $\mathbb{S}_{\rho,\sigma}$ on the $\rho-\sigma$ plane ($\tau=0$) with nonzero real values of both $\rho$ and $\sigma$, e.g. $(0, 0, 0, 0, \rho, \sigma, 0)$.
\item [$ \bullet $] The points $\mathbb{S}_{\rho,\tau}$ on the $\rho-\tau$ plane ($\sigma=0$) with nonzero real values of both $\rho$ and $\tau$, e.g. $(0, 0, 0, 0, 0, \rho, 0, \tau)$.
\item [$ \bullet $] The points $\mathbb{S}_{\sigma,\tau}$ on the $\sigma-\tau$ plane ($\rho=0$) with nonzero real values of both $\sigma$ and $\tau$, e.g. $(0, 0, 0, 0, 0, \sigma, \tau)$.
\item [$ \bullet $] The disconnected set of points $\mathbb{L}_{\rho}$ on a line with $\sigma$ and $\tau$ both being zero and $\rho$ being nonzero, e.g. $(0, 0, 0, 0, \rho, 0, 0)$.
\item [$ \bullet $] The disconnected set of points $\mathbb{L}_{\sigma}$ on a line with $\rho$ and $\tau$ both being zero and $\sigma$ being nonzero, e.g. $(0, 0, 0, 0, 0, \sigma, 0)$.
\item [$ \bullet $] The disconnected set of points $\mathbb{L}_{\tau}$ on a line with $\rho$ and $\sigma$ both being zero and $\tau$ being nonzero, e.g. $(0, 0, 0, 0, 0, 0, \tau)$.
\item [$ \bullet $] The origin $O$ of the underlying dual algebra space $\mathbb{R}^{7}$ with $\rho=\sigma=\tau=0$. The coordinate of $O$ is just $(0, 0, 0, 0, 0, 0, 0)$.
\end{itemize}

From the coadjoint action given by (\ref{coad-action-exprssn}), one finds all the coadjoint orbits associated with the above enumerated points in the embedded $\mathbb{R}^{3}$ space. These coadjoint orbits are listed below
\begin{itemize}
\item [$ \circ $] $\mathbb{R}_{4}$ spaces $\mathcal{O}^{\rho,\sigma,0}_{4}$ each of which passes through a point lying in a disconnected toplogical space $\mathbb{S}_{\rho,\sigma}$, with $\rho\neq 0$ and $\sigma\neq 0$ rendering to its disconnectedness in the usual Euclidean topology. A generic point on such an orbit (fixed $\rho$ and $\sigma$) is given by $(k_{1}, k_{2}, k_{3}, k_{4}, \rho, \sigma, 0)$ with each of $k_{1}$, $k_{2}$, $k_{3}$, and $k_{4}$ assuming real values.
\item [$ \circ $] $\mathbb{R}_{4}$ spaces $\mathcal{O}^{\rho, 0, \tau}_{4}$ each of which passes through a point lying in the disconnected set $\mathbb{S}_{\rho,\tau}$, where disconnectedness refers to one in the usual Euclidean topology with each of $\rho$ and $\tau$ being nonzero. A generic point on such an orbit (fixed $\rho$ and $\tau$) is given by $(k_{1}, k_{2}, k_{3}, k_{4}, \rho, 0, \tau)$ with each of $k_{1}$, $k_{2}$, $k_{3}$, and $k_{4}$ assuming real values.
\item [$ \circ $] $\mathbb{R}_{4}$ spaces $\mathcal{O}^{0, \sigma, \tau}_{4}$ each of which passes through a point
    lying in the disconnected set $\mathbb{S}_{\sigma,\tau}$, the disconnectedness in the usual Euclidean topology being attributed to $\sigma\neq 0$ and $\tau\neq 0$. A generic point on such an orbit (fixed $\sigma$ and $\tau$) is given by $(k_{1}, k_{2}, k_{3}, k_{4}, 0, \sigma, \tau)$ with each of $k_{1}$, $k_{2}$, $k_{3}$, and $k_{4}$ assuming real values.
\item [$ \circ $] $\mathbb{R}_{4}$ spaces $\mathcal{O}^{\rho, 0, 0}_{4}$ each of which passes through a point lying in the disconnected set $\mathbb{L}_{\rho}$ ($\rho\neq 0$ contributes to its disconnectedness in the usual Euclidean topology). A generic point on such an orbit (fixed $\rho$) is given by $(k_{1}, k_{2}, k_{3}, k_{4}, \rho, 0, 0)$ with each of $k_{1}$, $k_{2}$, $k_{3}$, and $k_{4}$ assuming real values.
\item [$ \circ $] $\mathbb{R}_{2}$-plane $^{c_{3},c_{4}}\mathcal{O}^{0, \sigma, 0}_{2}$ due to a fixed ordered pair $(c_{3},c_{4})$. Such a plane lies in the $\mathbb{R}^{4}$ space that passes through each point of $\mathbb{L}_{\sigma}$, where $\mathbb{L}_{\sigma}$ is the punctured line with $\sigma\neq 0$. A generic point on such an orbit (fixed $\sigma$) is given by $(k_{1}, k_{2}, c_{3}, c_{4}, 0, \sigma, 0)$ with both $k_{1}$ and $k_{2}$ assuming real values.
\item [$ \circ $] $\mathbb{R}_{2}$ plane $^{c_{1},c_{2}}\mathcal{O}^{0, 0, \tau}_{2}$ due to a fixed ordered pair $(c_{1},c_{2})$. The plane lies in the $\mathbb{R}^{4}$ space that passes through each point of the punctured line $\mathbb{L}_{\tau}$ with $\tau\neq 0$. A generic point on such an orbit (fixed $\tau$) is given by $(c_{1}, c_{2}, k_{3}, k_{4}, 0, 0, \tau)$ with both $k_{3}$ and $k_{4}$ assuming real values.
\item [$ \circ $] $0$-dimensional point $^{c_{1},c_{2},c_{3},c_{4}}\mathcal{O}^{0, 0, 0}_{0}$ due to a fixed ordered quadruple $(c_{1},c_{2},c_{3},c_{4})$. Such a point lies in the $\mathbb{R}^{4}$ space that passes through the origin $O$. The corresponding zero dimensional orbit is denoted as $(c_{1}, c_{2}, c_{3}, c_{4}, 0, 0, 0)$.
    \end{itemize}

    We are now all set to resume our computations on finding the unitary irreducible representations of $\g$. From the method of orbits (see \cite{Kirillovbook}), we know that the unitary irreducible representations of the connected simply connected nilpotent Lie group $\g$ are in $1$-$1$ correspondence with its coadjoint orbits. The UIRs corresponding to the $4$ dimensional orbits have functional dimension $2$, i.e. the representation space is $L^{2}(\mathbb{R}^{2})$ with respect to the usual Lebesgue measure. Let us compute the UIRs $U^{\rho}_{\sigma,\tau}$, $U^{\rho}_{\sigma,0}$, $U^{\rho}_{0,\tau}$, $U^{0}_{\sigma,\tau}$, and $U^{\rho}_{0,0}$ corresponding to the coadjoint orbits $\mathcal{O}^{\rho, \sigma, \tau}_{4}$, $\mathcal{O}^{\rho, \sigma, 0}_{4}$, $\mathcal{O}^{\rho, 0, \tau}_{4}$, $\mathcal{O}^{0, \sigma, \tau}_{4}$, and $\mathcal{O}^{\rho, 0, 0}_{4}$, respectively.

    The most crucial part for the remaining task is to find the polarizing subalgebra, i.e. a maximal subalgebra $\mathfrak{h}$ of $\G$ which is subordinate to $F\in\G^{*}$ with representation given by (\ref{dual-algbra-rep}). In other words, $\mathfrak{h}$ must satisfy $F|_{[\mathfrak{h},\mathfrak{h}]}=0$. For the $4$ dimensional orbits, the polarizing subalgebra has to have dimension equal to $\frac{7+3}{2}$, i.e. $5$. The maximal abelian subalgebra of $\G$ serves as the polarizing subalgebra in this case. A generic element $h$ of the corresponding abelian subgroup $H\subset\g$ has the following matrix representation:
    \begin{equation}\label{abln-subgrp}
    h(\theta,\phi,\psi,p_{1},q_{2})=\begin{bmatrix}1&0&0&-\frac{\alpha}{2}p_{1}&0&0&\frac{\alpha}{2}q_{2}&\theta\\0&1&0&0&0&0&\frac{\beta}{2}p_{1}&\phi\\0&0&1&-\frac{\gamma}{2}q_{2}&0&0&0&\psi\\0&0&0&1&0&0&0&0\\0&0&0&0&1&0&0&q_{2}\\0&0&0&0&0&1&0&p_{1}\\0&0&0&0&0&0&1&0\\0&0&0&0&0&0&0&1\end{bmatrix}.
    \end{equation}
    Let us consider the following section $\delta:H\backslash\g\rightarrow\g$. The matrix representation of the section $\delta$ then reads
    \begin{equation}\label{section-rep}
    \delta(r_{1},s_{2})=\begin{bmatrix}1&0&0&0&-\frac{\alpha}{2}s_{2}&\frac{\alpha}{2}r_{1}&0&0\\0&1&0&0&0&-\frac{\beta}{2}s_{2}&0&0\\0&0&1&0&\frac{\gamma}{2}r_{1}&0&0&0\\0&0&0&1&0&0&0&r_{1}\\0&0&0&0&1&0&0&0\\0&0&0&0&0&1&0&0\\0&0&0&0&0&0&1&s_{2}\\0&0&0&0&0&0&0&1\end{bmatrix}.
    \end{equation}
    All we have to do now is to solve the master equation,
    \begin{eqnarray}\label{master-eqn}
    \lefteqn{\left[\begin{smallmatrix}1&0&0&0&-\frac{\alpha}{2}s_{2}&\frac{\alpha}{2}r_{1}&0&0\\0&1&0&0&0&-\frac{\beta}{2}s_{2}&0&0\\0&0&1&0&\frac{\gamma}{2}r_{1}&0&0&0\\0&0&0&1&0&0&0&r_{1}\\0&0&0&0&1&0&0&0\\0&0&0&0&0&1&0&0\\0&0&0&0&0&0&1&s_{2}\\0&0&0&0&0&0&0&1\end{smallmatrix}\right]\left[\begin{smallmatrix}1&0&0&-\frac{\alpha}{2}p_{1}&-\frac{\alpha}{2}p_{2}&\frac{\alpha}{2}q_{1}&\frac{\alpha}{2}q_{2}&\theta\\0&1&0&0&0&-\frac{\beta}{2}p_{2}&\frac{\beta}{2}p_{1}&\phi\\0&0&1&-\frac{\gamma}{2}q_{2}&\frac{\gamma}{2}q_{1}&0&0&\psi\\0&0&0&1&0&0&0&q_{1}\\0&0&0&0&1&0&0&q_{2}\\0&0&0&0&0&1&0&p_{1}\\0&0&0&0&0&0&1&p_{2}\\0&0&0&0&0&0&0&1\end{smallmatrix}\right]}\nonumber\\
    &&=\left[\begin{smallmatrix}1&0&0&-\frac{\alpha}{2}p_{1}&-\frac{\alpha}{2}(p_{2}+s_{2})&\frac{\alpha}{2}(q_{1}+r_{1})&\frac{\alpha}{2}q_{2}&\theta-\frac{\alpha}{2}q_{2}s_{2}+\frac{\alpha}{2}p_{1}r_{1}\\0&1&0&0&0&-\frac{\beta}{2}(p_{2}+s_{2})&\frac{\beta}{2}p_{1}&\phi-\frac{\beta}{2}p_{1}s_{2}\\0&0&1&-\frac{\gamma}{2}q_{2}&\frac{\gamma}{2}(q_{1}+r_{1})&0&0&\psi+\frac{\gamma}{2}q_{2}r_{1}\\0&0&0&1&0&0&0&q_{1}+r_{1}\\0&0&0&0&1&0&0&q_{2}\\0&0&0&0&0&1&0&p_{1}\\0&0&0&0&0&0&1&p_{2}+s_{2}\\0&0&0&0&0&0&0&1\end{smallmatrix}\right]\label{master-eqn-compr}\\
    &&=\left[\begin{smallmatrix}1&0&0&-\frac{\alpha}{2}A&0&0&\frac{\alpha}{2}B&C\\0&1&0&0&0&0&\frac{\beta}{2}A&D\\0&0&1&-\frac{\gamma}{2}B&0&0&0&E\\0&0&0&1&0&0&0&0\\0&0&0&0&1&0&0&B\\0&0&0&0&0&1&0&A\\0&0&0&0&0&0&1&0\\0&0&0&0&0&0&0&1\end{smallmatrix}\right]\left[\begin{smallmatrix}1&0&0&0&-\frac{\alpha}{2}F&\frac{\alpha}{2}G&0&0\\0&1&0&0&0&-\frac{\beta}{2}F&0&0\\0&0&1&0&\frac{\gamma}{2}G&0&0&0\\0&0&0&1&0&0&0&G\\0&0&0&0&1&0&0&0\\0&0&0&0&0&1&0&0\\0&0&0&0&0&0&1&F\\0&0&0&0&0&0&0&1\end{smallmatrix}\right]\nonumber\\
    &&=\left[\begin{smallmatrix}1&0&0&-\frac{\alpha}{2}A&-\frac{\alpha}{2}F&\frac{\alpha}{2}G&\frac{\alpha}{2}B&C+\frac{\alpha}{2}BF-\frac{\alpha}{2}GA\\0&1&0&0&0&-\frac{\beta}{2}F&\frac{\beta}{2}A&D+\frac{\beta}{2}AF\\0&0&1&-\frac{\gamma}{2}B&\frac{\gamma}{2}G&0&0&E-\frac{\gamma}{2}BG\\0&0&0&1&0&0&0&G\\0&0&0&0&1&0&0&B\\0&0&0&0&0&1&0&A\\0&0&0&0&0&0&1&F\\0&0&0&0&0&0&0&1\end{smallmatrix}\right].\label{master-eqn-compr-othr}
    \end{eqnarray}
    The unknowns $A,B,C,D,E,F$ and $G$ can easily be computed by comparing (\ref{master-eqn-compr}) with (\ref{master-eqn-compr-othr}). We get
    \begin{equation}\label{unknowns}
    \begin{split}
    &A=p_{1},\;
    B=q_{2},\;
    G=r_{1}+q_{1},\;
    F=s_{2}+p_{2},\\
    &C=\theta-\alpha q_{2}s_{2}+\alpha p_{1}r_{1}+\frac{\alpha}{2}q_{1}p_{1}-\frac{\alpha}{2}q_{2}p_{2},\\
    &D=\phi-\beta p_{1}s_{2}-\frac{\beta}{2}p_{1}p_{2},\;
    E=\psi+\gamma q_{2}r_{1}+\frac{\gamma}{2}q_{1}q_{2}.
    \end{split}
    \end{equation}
    Now, the dual algebra vector lying in the $4$ dimensional coadjoint orbit $\mathcal{O}^{\rho,\sigma,\tau}_{4}$ was found to be $(0,0,0,0,\rho,\sigma,\tau)$. Using (\ref{unknowns}), a family of representations $U^{\rho}_{\sigma,\tau}$ associated with these coadjoint orbits follow immediately
    \begin{eqnarray}\label{nonzero-rep}
    \lefteqn{(U^{\rho}_{\sigma,\tau}(\theta,\phi,\psi,q_{1},q_{2},p_{1},p_{2})f)(r_{1},s_{2})}\nonumber\\
    &&=e^{i\rho(\theta-\alpha q_{2}s_{2}+\alpha p_{1}r_{1}+\frac{\alpha}{2}q_{1}p_{1}-\frac{\alpha}{2}q_{2}p_{2})}e^{i\sigma(\phi-\beta p_{1}s_{2}-\frac{\beta}{2}p_{1}p_{2})}\nonumber\\
    &&\times e^{i\tau(\psi+\gamma q_{2}r_{1}+\frac{\gamma}{2}q_{2}q_{1})}f(r_{1}+q_{1},s_{2}+p_{2}),
    \end{eqnarray}
    where none of $\rho$, $\sigma$ and $\tau$ are zero and $f\in L^{2}(\mathbb{R}^{2},dr_{1}ds_{2})$.

    It is worth mentioning in this context that the family of UIRs (\ref{nonzero-rep}) is computed in (\cite{ncqmjmp}), albeit using an $F\in\G^{*}$ different from one given by (\ref{dual-algbra-rep}).

    The required dimension of the polarizing subalgebra $\mathfrak{h}$ due to the other coadjoint orbits is also $5$.
    And hence, the polarizing subalgebra that was used to compute the UIRs associated with the orbits $\mathcal{O}^{\rho,\sigma,\tau}_{4}$, also serves for the other $4$ dimensional orbits of $\g$. Therefore, the results, obtained in (\ref{unknowns}), apply to all other $4$ dimensional coadjoint orbits, as well.

    Knowing that the $4$-dimensional orbit $\mathcal{O}^{\rho}_{\sigma,0}$ passes through the point $(0, 0, 0, 0, \rho, \sigma, 0)$, we can easily obtain the corresponding family of UIRs:
    \begin{eqnarray}\label{tau-zero-rep}
    \lefteqn{(U^{\rho}_{\sigma,0}(\theta,\phi,\psi,q_{1},q_{2},p_{1},p_{2})f)(r_{1},s_{2})}\nonumber\\
    &&=e^{i\rho(\theta-\alpha q_{2}s_{2}+\alpha p_{1}r_{1}+\frac{\alpha}{2}q_{1}p_{1}-\frac{\alpha}{2}q_{2}p_{2})}e^{i\sigma(\phi-\beta p_{1}s_{2}-\frac{\beta}{2}p_{1}p_{2})}f(r_{1}+q_{1},s_{2}+p_{2}),
    \end{eqnarray}
    where $f\in L^{2}(\mathbb{R}^{2},dr_{1}ds_{2})$.

    Now, the orbit $\mathcal{O}^{\rho}_{0,\tau}$ was found to pass through the point $(0, 0, 0, 0, \rho, 0, \tau)$ of the dual algebra space $\G^{*}$. Therefore, the continuous family of UIRs corresponding to these $4$-dimensional coadjoint orbits follows as
    \begin{eqnarray}\label{sigma-zero-rep}
    \lefteqn{(U^{\rho}_{0,\tau}(\theta,\phi,\psi,q_{1},q_{2},p_{1},p_{2})f)(r_{1},s_{2})}\nonumber\\
    &&=e^{i\rho(\theta-\alpha q_{2}s_{2}+\alpha p_{1}r_{1}+\frac{\alpha}{2}q_{1}p_{1}-\frac{\alpha}{2}q_{2}p_{2})}e^{i\tau(\psi+\gamma q_{2}r_{1}+\frac{\gamma}{2}q_{2}q_{1})}f(r_{1}+q_{1},s_{2}+p_{2}),
    \end{eqnarray}
    where $f\in L^{2}(\mathbb{R}^{2},dr_{1}ds_{2})$.

    Also, $\mathcal{O}^{0,\sigma,\tau}_{4}$, being a $4$ dimensional coadjoint orbit, passes through the point\\
    $(0, 0, 0, 0, 0, \sigma, \tau)\in\G^{*}$. Therefore, the corresponding family of UIRs is given by
    \begin{eqnarray}\label{rho-zero-rep}
    \lefteqn{(U^{0}_{\sigma,\tau}(\theta,\phi,\psi,q_{1},q_{2},p_{1},p_{2})f)(r_{1},s_{2})}\nonumber\\
    &&=e^{i\sigma(\phi-\beta p_{1}s_{2}-\frac{\beta}{2}p_{1}p_{2})}e^{i\tau(\psi+\gamma q_{2}r_{1}+\frac{\gamma}{2}q_{2}q_{1})}f(r_{1}+q_{1},s_{2}+p_{2}),
    \end{eqnarray}
    where both $\sigma$ and $\tau$ are nonzero and $f\in L^{2}(\mathbb{R}^{2},dr_{1}ds_{2})$.

    The only remaining $4$ dimensional coadjoint orbit is $\mathcal{O}^{\rho}_{0,0}$ which passes through the point $(0, 0, 0, 0, \rho, 0, 0)\in\G^{*}$. The family of unitary irreducible representations associated with these orbits are found to be
    \begin{eqnarray}\label{rho-nonzerrep}
    \lefteqn{(U^{\rho}_{0,0}(\theta,\phi,\psi,q_{1},q_{2},p_{1},p_{2})f)(r_{1},s_{2})}\nonumber\\
    &&=e^{i\rho(\theta-\alpha q_{2}s_{2}+\alpha p_{1}r_{1}+\frac{\alpha}{2}q_{1}p_{1}-\frac{\alpha}{2}q_{2}p_{2})}f(r_{1}+q_{1},s_{2}+p_{2}),
    \end{eqnarray}
    where $\rho$ is nonzero and $f\in L^{2}(\mathbb{R}^{2},dr_{1}ds_{2})$.

    There are three families of $2$-dimensional coadjoint orbits in the present setting. For any nonzero triple $(\rho,\sigma,\tau)$ with $\rho^{2}\alpha^{2}-\gamma\beta\sigma\tau=0$, we obtained a family of coadjoint orbits denoted by $^{\kappa,\delta}\mathcal{O}^{\rho,\zeta}_{2}$ where the ordered pair $(\kappa,\delta)\in\mathbb{R}^{2}$ and $\zeta\in(-\infty,0)\cup(0,\infty)$. These orbits pass through the points $(0,0,\kappa,\delta,\rho,\frac{\rho}{\zeta},\frac{\zeta\alpha^{2}\rho}{\gamma\beta})\in\G^{*}$. The other $2$-dimensional orbits $^{c_{1},c_{2}}\mathcal{O}^{0, 0, \tau}_{2}$ and $^{c_{3}, c_{4}}\mathcal{O}^{0, \sigma, 0}_{2}$ pass through the points $(c_{1}, c_{2}, 0, 0, 0, 0, \tau)$ and $(0, 0, c_{3}, c_{4}, 0, \sigma, 0)$ of $\G^{*}$, respectively. For the $2$-dimensional case, the required dimension of the polarizing subalgebra is no longer $5$. It is now $\frac{7+5}{2}=6$.

    Let us, first, compute the unitary irreducible representations associated with the $2$-dimensional coadjoint orbits $^{c_{1},c_{2}}\mathcal{O}^{0, 0, \tau}_{2}$ and $^{c_{3}, c_{4}}\mathcal{O}^{0, \sigma, 0}_{2}$. In case of the $2$-dimensional coadjoint orbits $^{c_{1},c_{2}}\mathcal{O}^{0, 0, \tau}_{2}$, a generic element of the polarizing subalgebra $\mathfrak{h}$ is given by
    $$\begin{bmatrix}0&0&0&-\frac{\alpha}{2}x^{1}&-\frac{\alpha}{2}x^{2}&\frac{\alpha}{2}x^{3}&0&x^{5}\\0&0&0&0&0&-\frac{\beta}{2}x^{2}&\frac{\beta}{2}x^{1}&x^{6}\\0&0&0&0&\frac{\gamma}{2}x^{3}&0&0&x^{7}\\0&0&0&0&0&0&0&x^{3}\\0&0&0&0&0&0&0&0\\0&0&0&0&0&0&0&x^{1}\\0&0&0&0&0&0&0&x^{2}\\0&0&0&0&0&0&0&0\end{bmatrix}.
    $$
    One can easily verify that under the above choice of polarizing subalgebra the following holds
    \begin{equation}
    F|_{[\mathfrak{h},\mathfrak{h}]}=0,
    \end{equation}
    where the matrix representation of a dual algebra element $F$ is given by (\ref{dual-algbra-rep}). Therefore, an element of the corresponding subgroup $H\subset\g$ (note that this subgroup is no longer abelian) is as follows
    \begin{equation}\label{pol-subalgbr-nonabln}
    h(\theta,\phi,\psi,p_{1},p_{2},q_{1})=\begin{bmatrix}1&0&0&-\frac{\alpha}{2}p_{1}&-\frac{\alpha}{2}p_{2}&\frac{\alpha}{2}q_{1}&0&\theta\\0&1&0&0&0&-\frac{\beta}{2}p_{2}&\frac{\beta}{2}p_{1}&\phi\\0&0&1&0&\frac{\gamma}{2}q_{1}&0&0&\psi\\0&0&0&1&0&0&0&q_{1}\\0&0&0&0&1&0&0&0\\0&0&0&0&0&1&0&p_{1}\\0&0&0&0&0&0&1&p_{2}\\0&0&0&0&0&0&0&1\end{bmatrix}.
    \end{equation}
    We now consider the following section $\delta:H\backslash\g\rightarrow\g$ given by
    \begin{equation}\label{section-two-dim}
    \delta(r)=\begin{bmatrix}1&0&0&0&0&0&\frac{\alpha}{2}r&0\\0&1&0&0&0&0&0&0\\0&0&1&-\frac{\gamma}{2}r&0&0&0&0\\0&0&0&1&0&0&0&0\\0&0&0&0&1&0&0&r\\0&0&0&0&0&1&0&0\\0&0&0&0&0&0&1&0\\0&0&0&0&0&0&0&1\end{bmatrix}.
    \end{equation}
    Then the corresponding master equation leads to
    \begin{eqnarray}\label{master-eqn-two-dim}
    \lefteqn{\left[\begin{smallmatrix}1&0&0&0&0&0&\frac{\alpha}{2}r&0\\0&1&0&0&0&0&0&0\\0&0&1&-\frac{\gamma}{2}r&0&0&0&0\\0&0&0&1&0&0&0&0\\0&0&0&0&1&0&0&r\\0&0&0&0&0&1&0&0\\0&0&0&0&0&0&1&0\\0&0&0&0&0&0&0&1\end{smallmatrix}\right]\left[\begin{smallmatrix}1&0&0&-\frac{\alpha}{2}p_{1}&-\frac{\alpha}{2}p_{2}&\frac{\alpha}{2}q_{1}&\frac{\alpha}{2}q_{2}&\theta\\0&1&0&0&0&-\frac{\beta}{2}p_{2}&\frac{\beta}{2}p_{1}&\phi\\0&0&1&-\frac{\gamma}{2}q_{2}&\frac{\gamma}{2}q_{1}&0&0&\psi\\0&0&0&1&0&0&0&q_{1}\\0&0&0&0&1&0&0&q_{2}\\0&0&0&0&0&1&0&p_{1}\\0&0&0&0&0&0&1&p_{2}\\0&0&0&0&0&0&0&1\end{smallmatrix}\right]}\nonumber\\
    &&=\left[\begin{smallmatrix}1&0&0&-\frac{\alpha}{2}p_{1}&-\frac{\alpha}{2}p_{2}&\frac{\alpha}{2}q_{1}&\frac{\alpha}{2}(q_{2}+r)&\theta+\frac{\alpha}{2}p_{2}r\\0&1&0&0&0&-\frac{\beta}{2}p_{2}&\frac{\beta}{2}p_{1}&\phi\\0&0&1&-\frac{\gamma}{2}(q_{2}+r)&\frac{\gamma}{2}q_{1}&0&0&\psi-\frac{\gamma}{2}q_{1}r\\0&0&0&1&0&0&0&q_{1}\\0&0&0&0&1&0&0&q_{2}+r\\0&0&0&0&0&1&0&p_{1}\\0&0&0&0&0&0&1&p_{2}\\0&0&0&0&0&0&0&1\end{smallmatrix}\right]\label{master-eqn-compr-two-dim}\\
    &&=\left[\begin{smallmatrix}1&0&0&-\frac{\alpha}{2}A&-\frac{\alpha}{2}B&\frac{\alpha}{2}C&0&D\\0&1&0&0&0&-\frac{\beta}{2}B&\frac{\beta}{2}A&E\\0&0&1&0&\frac{\gamma}{2}C&0&0&F\\0&0&0&1&0&0&0&C\\0&0&0&0&1&0&0&0\\0&0&0&0&0&1&0&A\\0&0&0&0&0&0&1&B\\0&0&0&0&0&0&0&1\end{smallmatrix}\right]\left[\begin{smallmatrix}1&0&0&0&0&0&\frac{\alpha}{2}G&0\\0&1&0&0&0&0&0&0\\0&0&1&-\frac{\gamma}{2}G&0&0&0&0\\0&0&0&1&0&0&0&0\\0&0&0&0&1&0&0&G\\0&0&0&0&0&1&0&0\\0&0&0&0&0&0&1&0\\0&0&0&0&0&0&0&1\end{smallmatrix}\right]\nonumber\\
    &&=\left[\begin{smallmatrix}1&0&0&-\frac{\alpha}{2}A&-\frac{\alpha}{2}B&\frac{\alpha}{2}C&\frac{\alpha}{2}G&D-\frac{\alpha}{2}BG\\0&1&0&0&0&-\frac{\beta}{2}B&\frac{\beta}{2}A&E\\0&0&1&-\frac{\gamma}{2}G&\frac{\gamma}{2}C&0&0&F+\frac{\gamma}{2}CG\\0&0&0&1&0&0&0&C\\0&0&0&0&1&0&0&G\\0&0&0&0&0&1&0&A\\0&0&0&0&0&0&1&B\\0&0&0&0&0&0&0&1\end{smallmatrix}\right].\label{master-eqn-compr-othr-two-dim}
    \end{eqnarray}
    The unknowns $A,B,C,D,E,F$ and $G$ can easily be computed by comparing (\ref{master-eqn-compr-two-dim}) with (\ref{master-eqn-compr-othr-two-dim}). We get
    \begin{equation}\label{unknowns-two-dim}
    \begin{split}
    &A=p_{1},\;
    B=p_{2},\;
    C=q_{1},\;
    G=q_{2}+r,\\
    &D=\theta+\alpha p_{2}r+\frac{\alpha}{2}p_{2}q_{2},\\
    &E=\phi,\;
    F=\psi-\gamma q_{1}r-\frac{\gamma}{2}q_{1}q_{2}.
    \end{split}
    \end{equation}
    Now, a dual algebra vector lying in the $2$ dimensional coadjoint orbit $^{c_{1}, c_{2}}\mathcal{O}^{0, 0, \tau}_{2}$ was found to be $(c_{1}, c_{2}, 0, 0, 0, 0, \tau)$. Using (\ref{unknowns-two-dim}), a family of representations $U^{c_{1}, c_{2}}_{0, 0,\tau}$ associated with these coadjoint orbits, for fixed $c_{1}$ and $c_{2}$, follow immediately
    \begin{eqnarray}\label{taunonzero-rep-two-dim}
    \lefteqn{(U^{c_{1}, c_{2}}_{0, 0, \tau}(\theta,\phi,\psi,q_{1},q_{2},p_{1},p_{2})f)(r)}\nonumber\\
    &&=e^{ic_{1}p_{1}+ic_{2}p_{2}}e^{i\tau(\psi-\gamma q_{1}r-\frac{\gamma}{2}q_{1}q_{2})}f(r+q_{2}),
    \end{eqnarray}
    where $\tau$ is nonzero and $f\in L^{2}(\mathbb{R},dr)$.

    Following exactly the same computations of the $2$ dimensional coadjoint orbits $^{c_{1}, c_{2}}\mathcal{O}^{0, 0, \tau}_{2}$ except for a different choice of $6$ dimensional polarizing subalgebra, one can derive the UIRs associated with the $2$-dimensional coadjoint orbits $^{c_{3}, c_{4}}\mathcal{O}^{0, \sigma, 0}_{2}$ with a fixed ordered pair $(c_{3},c_{4})$,
    \begin{eqnarray}\label{sigmanonzero-rep-two-dim}
    \lefteqn{(U^{c_{3}, c_{4}}_{0, \sigma, 0}(\theta,\phi,\psi,q_{1},q_{2},p_{1},p_{2})f)(s)}\nonumber\\
    &&=e^{ic_{3}q_{1}+ic_{4}q_{2}}e^{i\sigma(\phi-\beta p_{1}s-\frac{\beta}{2}p_{1}p_{2})}f(s+p_{2}),
    \end{eqnarray}
    where $\sigma$ is nonzero and $f\in L^{2}(\mathbb{R},ds)$.

Let us now compute the UIRs associated with the $2$-dimensional coadjoint orbits $^{\kappa,\delta}\mathcal{O}^{\rho,\zeta}_{2}$. We have already noted that any point on the surface $\mathbb{S}^{\rho,\zeta}$, in figure \ref{fig:figfirst}, is denoted by its coordinate $(\rho,\zeta)$ with $\rho\neq 0$ and $\zeta\in(-\infty,0)\cup(0,\infty)$. As has already been noted that each such $2$-dimensional coadjoint orbit $^{\kappa,\delta}\mathcal{O}^{\rho,\zeta}_{2}$ passes through the point $(0,0,\kappa,\delta,\rho,\frac{\rho}{\zeta},\frac{\zeta\alpha^{2}\rho}{\gamma\beta})\in\G^{*}$. The subgroup corresponding  to an astute choice of $6$-dimensional poralizing subalgebra is obtained by inserting $p_{1}=-\frac{\zeta\alpha q_{2}}{\beta}$ in (\ref{mtrx-relztn}). Now, repeating the same procedure of constructing the so-called master equation, one obtains the following family of unitary irreducible representations of $\g$ associated with the $2$-dimensional coadjoint orbits $^{\kappa,\delta}\mathcal{O}^{\rho,\zeta}_{2}$:
\begin{eqnarray}\label{nonzero-rep-two-dimensional}
    \lefteqn{(U^{\kappa,\delta}_{\rho,\zeta}(\theta,\phi,\psi,q_{1},q_{2},p_{1},p_{2})f)(s)}\nonumber\\
    &&=e^{i\kappa q_{1}+i\delta q_{2}+i\rho(\theta-\alpha q_{1}s-\frac{\alpha q_{1}p_{1}}{2}-\frac{\zeta\alpha^{2}q_{1}q_{2}}{2\beta})}e^{\frac{i\rho}{\zeta}(\phi+\beta p_{2}s+\frac{\zeta\alpha q_{2}p_{2}}{2}+\frac{\beta}{2}p_{1}p_{2})}\nonumber\\
    &&\times e^{i\frac{\zeta\rho\alpha^{2}}{\gamma\beta}\psi}f(s+p_{1}+\frac{\zeta\alpha q_{2}}{\beta}),
\end{eqnarray}
where $f\in L^{2}(\mathbb{R},ds)$. It is worth noting in this context that for each ordered pair $(\kappa,\delta)\in\mathbb{R}^{2}$ along with $\rho\neq 0$ and $\zeta\in(-\infty,0)\cup(0,\infty)$ satisfying $\rho=\sigma\zeta=\frac{\gamma\beta\tau}{\zeta\alpha^{2}}$, we obtain one unitary irreducible representation $U^{\kappa,\delta}_{\rho,\zeta}$ of $\g$.

    There is only one zero dimensional orbit for $\G$. The zero dimensional coadjoint orbit \\$^{c_{1}, c_{2}, c_{3}, c_{4}}\mathcal{O}^{0, 0, 0}_{0}$ passes through the point $(c_{1}, c_{2}, c_{3}, c_{4}, 0, 0, 0)$ of the dual Lie algebra $\G^{*}$. And hence, follows the associated family of $1$-dimensional representations,
    \begin{eqnarray}\label{allzero-rep-zero-dim}
    \lefteqn{U^{c_{1}, c_{2}, c_{3}, c_{4}}_{0, 0, 0}(\theta,\phi,\psi,q_{1},q_{2},p_{1},p_{2})}\nonumber\\
    &&=e^{ic_{1}p_{1}+ic_{2}p_{2}+ic_{3}q_{1}+ic_{4}q_{2}}.
    \end{eqnarray}

    \section{Representation of the Lie algebra \texorpdfstring{$\G$}{gNC}}\label{sec:lie-algbra-rep}
    The basis elements for the Lie algebra $\G$ are $Q_{1}$, $Q_{2}$, $P_{1}$, $P_{2}$, $\Theta$, $\Phi$, and $\Psi$, where the last three elements form the $3$ dimensional center of the underlying Lie algebra. One has to represent these basis elements as appropriate operators on the corresponding group representation space, (see section \ref{sec:coad-orbts}). We compute the various unitary irreducible group representations restricted to one-parameter subgroups and thereby find the Hilbert space operators associated with the respective group parameters using the following equation:
    \begin{equation}\label{generator-formula}
    \hat{X}_{\eta}=-iC\frac{dU(\eta)}{d\eta}\mid_{\eta=0},
    \end{equation}
    where $\eta$ is one of the group parameters of $\g$ and $C$ is a constant fixed by the corresponding UIR with appropriate dimension.

    We consider the following cases:

    \subsection{Case \texorpdfstring{$\rho\neq 0, \sigma\neq 0, \tau\neq 0$ with $\rho^{2}\alpha^{2}-\gamma\beta\sigma\tau\neq 0.$}{rho sigma tau Nonzero}}\label{subsec:all-nonzero}
    A family of unitary irreducible representations $U^{\rho}_{\sigma,\tau}$ associated with the $4$ dimensional coadjoint orbits $\mathcal{O}^{\rho,\sigma,\tau}_{4}$ were found (see (\ref{nonzero-rep})) in section \ref{sec:coad-orbts}. These representations are labeled by nonzero real values of $\rho$, $\sigma$, and $\tau$.
    Let us now consider a unitary operator $T$ on $L^{2}(\mathbb{R}^{2}, dr_{1}ds_{2})$ given by
    \begin{equation}\label{dilation-operator}
    (Tf)(r_{1},s_{2})=f(r_{1}, -s_{2}),
    \end{equation}
    with $f\in L^{2}(\mathbb{R}^{2},dr_{1}ds_{2})$. The inverse $T^{-1}$ turns out immediately to be equal to $T$.

    Then a very straightforward computation shows that
    \begin{equation}\label{unitary equivalence}
    T^{-1}U^{\rho}_{\sigma,\tau}T=\tilde{U}^{\rho}_{\sigma,\tau},
    \end{equation}
    with $f$ lying in $L^{2}(\mathbb{R}^{2},dr_{1}ds_{2})$ and $\tilde{U}^{\rho}_{\sigma,\tau}$ given as
    \begin{eqnarray}\label{nonzero-equivalent-rep}
    \lefteqn{(\tilde{U}^{\rho}_{\sigma,\tau}(\theta,\phi,\psi,q_{1},q_{2},p_{1},p_{2})f)(r_{1},s_{2})}\nonumber\\
    &&=e^{i\rho(\theta+\alpha q_{2}s_{2}+\alpha p_{1}r_{1}+\frac{\alpha}{2}q_{1}p_{1}-\frac{\alpha}{2}q_{2}p_{2})}e^{i\sigma(\phi+\beta p_{1}s_{2}-\frac{\beta}{2}p_{1}p_{2})}\nonumber\\
    &&\times e^{i\tau(\psi+\gamma q_{2}r_{1}+\frac{\gamma}{2}q_{2}q_{1})}f(r_{1}+q_{1},s_{2}-p_{2}).
    \end{eqnarray}

    Let us now take the inverse Fourier transform of (\ref{nonzero-equivalent-rep}) with respect to second coordinate $s_{2}$ and call it $r_{2}$. Then using (\ref{generator-formula}) with $C=\frac{1}{\rho\alpha}$ the noncentral elements of $\G$ can be represented as the following operators on $L^{2}(\mathbb{R}^{2}, dr_{1}dr_{2})$:
    \begin{equation}\label{algbr-rep-equivalent-nonzero}
    \begin{split}
    &\hat{Q}_{1}=r_{1}+i\vartheta\frac{\partial}{\partial r_{2}},\qquad
    \hat{Q}_{2}=r_{2},\\
    &\hat{P}_{1}=-i\hbar\frac{\partial}{\partial r_{1}},\qquad
    \hat{P}_{2}=-\frac{\bm{\mathcal{B}}}{\hbar}r_{1}-i\hbar\frac{\partial}{\partial r_{2}},
    \end{split}
    \end{equation}
    with the following identification:
    \begin{equation}\label{idntfctn-nonzero-rep}
    \hbar=\frac{1}{\rho\alpha},\;\;\vartheta=-\frac{\sigma\beta}{(\rho\alpha)^{2}}\;\hbox{and}\;\bm{\mathcal{B}}=-\frac{\tau\gamma}{(\rho\alpha)^{2}}.
    \end{equation}
    $B:=\frac{\bm{\mathcal{B}}}{\hbar}$, here, can be interpreted as the constant magnetic field applied normally to the $\hat{Q}_{1}\hat{Q}_{2}$-plane. A charged particle constrained to move on this $\hat{Q}_{1}\hat{Q}_{2}$-plane then gives rise to kinematical momenta $\hat{P}_{1}$ and $\hat{P}_{2}$. These gauge invariant momenta differ from the standard quantum mechanical ones by a factor of the underlying vector potentials. The factor involves the charge and the velocity of light which can conveniently be assumed as being equal to 1. These kinematical momenta, when expressed in terms of the relevant vector potential, satisfy the commutation relations as provided by (\ref{commutators-rep-equivalent-nonzero}) if $B$ is interpreted as the constant normal magnetic field applied to the $\hat{Q}_{1}\hat{Q}_{2}$-plane (see \cite{delducetalt}). These commutation relations between the relevant Hilbert space operators (\ref{algbr-rep-equivalent-nonzero}) read
    \begin{equation}\label{commutators-rep-equivalent-nonzero}
    \begin{split}
    &[\hat{Q}_{1},\hat{P}_{1}]=[\hat{Q}_{2},\hat{P}_{2}]=i\hbar\mathbb{I},\;\;
    [\hat{Q}_{1},\hat{Q}_{2}]=i\vartheta \mathbb{I},\\
    &[\hat{P}_{1},\hat{P}_{2}]=i\hbar B\mathbb{I},\;\;
    [\hat{Q}_{1},\hat{P}_{2}]=[\hat{Q}_{2},P_{1}]=0,
    \end{split}
    \end{equation}
    $\mathbb{I}$ being the identity operator on $L^{2}(\mathbb{R}^{2},dr_{1}dr_{2})$.

    In case of the constant external magnetic field acting perpendicular to the $\hat{Q}_{1}\hat{Q}_{2}$-plane, the magnetic field $B$, in terms of the vector potential $\vec{A}=(A_{1},A_{2})$, reads
    \begin{equation}\label{mag-vec-eq}
    B=\partial_{1}A_{2}-\partial_{2}A_{1}.
    \end{equation}

    Using (\ref{algbr-rep-equivalent-nonzero}) and the fact that $\hat{P}_{j}=-i\hbar\frac{\partial}{\partial r_{j}}-A_{j}$ with $j=1,2$ (the charge and the velocity of light are all taken to be unity), one immediately finds that $\vec{A}=(0,\frac{\bm{\mathcal{B}}}{\hbar}r_{1})=(0,Br_{1})$. This gauge is known as the {\em Landau gauge} in the physical literature (see \cite{delducetalt} and many articles cited therein). It is worth mentioning in this context that as $\hbar^{2}-\bm{\mathcal{B}}\vartheta\rightarrow 0$, the representation, to which (\ref{algbr-rep-equivalent-nonzero}) approaches, is a reducible representation since in that limiting situation, $\hat{P}_{2}+\frac{\bm{\mathcal{B}}}{\hbar}\hat{Q}_{1}\rightarrow 0$, i.e. $\hat{Q}_{1}$ becomes proportional to $\hat{P}_{2}$. One can verify that, indeed, the linear space formed by the functions of the form $\psi(r_{1},r_{2})=f(r_{2})e^{-i\left(\lambda-\frac{r_{1}}{\vartheta}\right)r_{2}}$ remain invariant under the action of the operators given by (\ref{algbr-rep-equivalent-nonzero}).
However, there exists a family of unitary irreducible representations of the Lie algebra $\G$ for $\hbar^{2}-\bm{\mathcal{B}}\vartheta=0$ as will be computed in the following subsection.

\subsection{Case \texorpdfstring{$\rho\neq 0, \sigma\neq 0, \tau\neq 0$ with $\rho^{2}\alpha^{2}-\gamma\beta\sigma\tau=0.$}{rho sigma tau Nonzero det Zero}}\label{subsec:all-nonzero-det-zero}
    Let us, now, consider the group representations $U^{\kappa,\delta}_{\rho,\zeta}$ (see \ref{nonzero-rep-two-dimensional}) associated with the $2$-dimensional coadjoint orbits $^{\kappa,\delta}\mathcal{O}^{\rho,\zeta}_{2}$. Taking the inverse Fourier transform of (\ref{nonzero-rep-two-dimensional}) and subsequent application of (\ref{generator-formula}), then, yields the configuration space representations of $\G$ on $L^{2}(\mathbb{R},dr)$:
\begin{equation}\label{rep-nonzero-det-zero}
\begin{aligned}
&\hat{Q}_{1}=-r,\quad
\hat{Q}_{2}=i\vartheta\frac{\partial}{\partial r},\\
&\hat{P}_{1}=\hbar\kappa+i\hbar\frac{\partial}{\partial r},\quad
\hat{P}_{2}=\hbar\delta+\frac{\hbar r}{\vartheta},
\end{aligned}
\end{equation}
with the following identification:
\begin{equation}\label{idntfctn-nonzero-rep-zero-det}
\hbar=\frac{1}{\rho\alpha}\quad\hbox{and}\quad\vartheta=-\frac{\sigma\beta}{(\rho\alpha)^{2}}.
\end{equation}

The commutators between the relevant Hilbert space operators (\ref{rep-nonzero-det-zero}) now read
\begin{equation}\label{commutators-rep-nonzero-zero-det}
\begin{split}
&[\hat{Q}_{1},\hat{P}_{1}]=[\hat{Q}_{2},\hat{P}_{2}]=i\hbar\mathbb{I},\;\;
 [\hat{Q}_{1},\hat{Q}_{2}]=i\vartheta \mathbb{I},\\
&[\hat{P}_{1},\hat{P}_{2}]=\frac{i\hbar^{2}}{\vartheta}\mathbb{I},\;\;
 [\hat{Q}_{1},\hat{P}_{2}]=[\hat{Q}_{2},P_{1}]=0,
\end{split}
\end{equation}
$\mathbb{I}$ being the identity operator on $L^{2}(\mathbb{R},dr)$. It is easy to see that if in addition to (\ref{idntfctn-nonzero-rep-zero-det}), we take $\bm{\mathcal{B}}=-\frac{\tau\gamma}{(\rho\alpha)^{2}}$, then $\rho^{2}\alpha^{2}-\gamma\beta\sigma\tau=0$ translates to $\hbar^{2}-\bm{\mathcal{B}}\vartheta=0$ which is exactly what we are looking at.

    \subsection{Case \texorpdfstring{$\rho\neq 0, \sigma\neq 0, \tau=0.$}{rho sigma Nonzero}}\label{subsec:rho-sigma-nonzero}
    Let us consider the group representations $U^{\rho}_{\sigma, 0}$, pertaining to the $4$ dimensional coadjoint orbits $\mathcal{O}^{\rho, \sigma, 0}_{4}$ and given by equation (\ref{tau-zero-rep}). We can find a unitary operator on $L^{2}(\mathbb{R}^{2}, dr_{1}ds_{2})$ to obtain a unitary irreducible representation which will be equivalent to the one given by (\ref{tau-zero-rep}). One then has to take the inverse Fourier transform of the equivalent representation thus obtained with respect to the second coordinate. The pertinent representation of the algebra then reads off using (\ref{generator-formula}) with $C=\frac{1}{\rho\alpha}$
    \begin{equation}\label{algbr-rep-rho-sigma-nonzero}
    \begin{split}
    &\hat{Q}_{1}=r_{1}+i\vartheta\frac{\partial}{\partial r_{2}},\;\;
    \hat{Q}_{2}=r_{2},\\
    &\hat{P}_{1}=-i\hbar\frac{\partial}{\partial r_{1}},\;\;
    \hat{P}_{2}=-i\hbar\frac{\partial}{\partial r_{2}},
    \end{split}
    \end{equation}
    with the same identification given by (\ref{idntfctn-nonzero-rep}).
    And the commutators between the corresponding Hilbert space operators read
    \begin{equation}\label{commutators-rep-rho-sigma-nonzero}
    \begin{split}
    &[\hat{Q}_{1},\hat{P}_{1}]=[\hat{Q}_{2},\hat{P}_{2}]=i\hbar\mathbb{I},\;\;
    [\hat{Q}_{1},\hat{Q}_{2}]=i\vartheta\mathbb{I},\\
    &[\hat{P}_{1},\hat{P}_{2}]=0,\;\;
    [\hat{Q}_{1},\hat{P}_{2}]=[\hat{Q}_{2},P_{1}]=0.
    \end{split}
    \end{equation}
    Note that here $\bm{\mathcal{B}}=-\frac{\tau\gamma}{(\rho\alpha)^{2}}=0$. Physically, it refers to the same system (\ref{commutators-rep-equivalent-nonzero}) with the magnetic field turned off.

    \subsection{Case \texorpdfstring{$\rho\neq 0, \sigma= 0, \tau\neq 0.$}{rho tau Nonzero}}\label{subsec:rho-tau-nonzero}
    A continuous family of unitary irreducible representations was found for the group $\g$ in (\ref{sigma-zero-rep}) arising from the coadjoint orbits $\mathcal{O}^{\rho, 0, \tau}_{4}$. We can now carry out a procedure similar to the one adopted in Sec (\ref{subsec:all-nonzero}) to find a unitary irreducible representation equivalent to (\ref{sigma-zero-rep}). The corresponding representation of the Lie algebra $\G$ then reads
    \begin{equation}\label{algbr-rep-rho-tau-nonzero}
    \begin{split}
    &\hat{Q}_{1}= r_{1},\;\;
    \hat{Q}_{2}=r_{2},\\
    &\hat{P}_{1}=-i\hbar\frac{\partial}{\partial r_{1}},\;\;
    \hat{P}_{2}=-\frac{\bm{\mathcal{B}}}{\hbar}r_{1}-i\hbar\frac{\partial}{\partial r_{2}},
    \end{split}
    \end{equation}
    where $\hbar$ and $\bm{\mathcal{B}}$ follow from (\ref{idntfctn-nonzero-rep}). Here also we have used $C=\frac{1}{\rho\alpha}$ in (\ref{generator-formula}) to compute the relevant noncentral generators. And the corresponding commutators are given as
    \begin{equation}\label{commutators-rep-rho-tau-nonzero}
    \begin{split}
    &[\hat{Q}_{1},\hat{P}_{1}]=[\hat{Q}_{2},\hat{P}_{2}]=i\hbar\mathbb{I},\;\;
    [\hat{Q}_{1},\hat{Q}_{2}]=0,\\
    &[\hat{P}_{1},\hat{P}_{2}]=i\bm{\mathcal{B}}\mathbb{I},\;\;
    [\hat{Q}_{1},\hat{P}_{2}]=[\hat{Q}_{2},P_{1}]=0.
    \end{split}
    \end{equation}
    Physically, (\ref{commutators-rep-rho-tau-nonzero}) just represents a {\em Landau system} in the presence of a constant magnetic field $B=\frac{\bm{\mathcal{B}}}{\hbar}$.

    \subsection{Case \texorpdfstring{$\rho\neq0, \sigma= 0, \tau= 0.$}{rho Nonzero}}\label{subsec:rho-nonzero}
     In this case as well, we obtain an irreducible representation of $\g$, unitarily equivalent to $U^{\rho}_{0,0}$ given by (\ref{rho-nonzerrep}). The associated $4$ dimensional coadjoint orbits were $\mathcal{O}^{\rho, 0, 0}_{4}$. Now the representation of $\G$ on $L^{2}(\mathbb{R}^{2},dr_{1}dr_{2})$ reads off
    \begin{equation}\label{algbr-rep-rho-nonzero}
    \begin{split}
    &\hat{Q}_{1}=r_{1},\;\;
    \hat{Q}_{2}=r_{2},\\
    &\hat{P}_{1}=-i\hbar\frac{\partial}{\partial r_{1}},\;\;
    \hat{P}_{2}=-i\hbar\frac{\partial}{\partial r_{2}},
    \end{split}
    \end{equation}
    with the canonical commutation relations of standard quantum mechanics given by
    \begin{equation}\label{commutators-rho-nonzero}
    \begin{split}
    &[\hat{Q}_{1},\hat{P}_{1}]=[\hat{Q}_{2},\hat{P}_{2}]=i\hbar\mathbb{I},\;\;
    [\hat{Q}_{1},\hat{Q}_{2}]=0,\\
    &[\hat{P}_{1},\hat{P}_{2}]=0,\;\;
    [\hat{Q}_{1},\hat{P}_{2}]=[\hat{Q}_{2},P_{1}]=0.
    \end{split}
    \end{equation}
    We, therefore, find that the unitary irreducible representation of standard quantum mechanics is sitting inside the unitary dual of the triply extended group $\g$ of translations in $\mathbb{R}^{4}$.

    \subsection{Case \texorpdfstring{$\rho=0, \sigma\neq 0, \tau\neq 0.$}{sigma tau Nonzero}}\label{subsec:sigma-tau-nonzero}
      A family of unitary irreducible representations of $\g$ equivalent to $U^{0}_{\sigma, \tau}$ (see \ref{rho-zero-rep}), corresponding to the $4$ dimensional coadjoint orbits $\mathcal{O}^{0, \sigma, \tau}_{4}$, has the following Lie algebra representation on $L^{2}(\mathbb{R}^{2}, dr_{1}dr_{2})$:
     \begin{equation}\label{algbr-rep-sigma-tau-nonzero}
    \begin{split}
    &\hat{Q}_{1}=i\kappa_{1}\frac{\partial}{\partial r_{2}},\;\;
    \hat{Q}_{2}=r_{2},\\
    &\hat{P}_{1}=-i\frac{\partial}{\partial r_{1}},\;\;
    \hat{P}_{2}=-\kappa_{2}r_{1},
    \end{split}
    \end{equation}
    with $\kappa_{1}=-\sigma\beta$ and $\kappa_{2}=-\tau\gamma$.
    The corresponding commutators read
    \begin{equation}\label{commutators-rep-sigma-tau-nonzero}
    \begin{split}
    &[\hat{Q}_{1},\hat{P}_{1}]=[\hat{Q}_{2},\hat{P}_{2}]=0,\;\;
    [\hat{Q}_{1},\hat{Q}_{2}]=i\kappa_{1}\mathbb{I},\\
    &[\hat{P}_{1},\hat{P}_{2}]=i\kappa_{2}\mathbb{I},\;\;
    [\hat{Q}_{1},\hat{P}_{2}]=[\hat{Q}_{2},P_{1}]=0.
    \end{split}
    \end{equation}
    (\ref{commutators-rep-sigma-tau-nonzero}) could be considered to represent an uncoupled system of two noncommutative planes. Referring back to (\ref{rho-zero-rep}), the   $\hat{Q}_{1}$, $\hat{Q}_{2}$, $\hat{P}_{1}$, and $\hat{P}_{2}$ in  (\ref{algbr-rep-sigma-tau-nonzero})are just the generators corresponding to abstract group parameters $p_{1}$, $p_{2}$, $q_{1}$, and $q_{2}$, respectively that represent translations in $\mathbb{R}^{4}$. They are not to be treated as position or momentum variables as they were in all four preceding cases and hence they are taken to be all dimensionless. The absence of $\hbar$ in the representation (\ref{algbr-rep-sigma-tau-nonzero}) of the noncentral generators also indicates the uncoupledness between the two underlying noncommutative planes.

    \subsection{Case \texorpdfstring{$\rho= 0, \sigma= 0, \tau\neq 0.$}{tau Nonzero}}\label{subsec:tau-nonzero}
    This situation is very much similar to that of (\ref{subsec:sigma-tau-nonzero}) except that we have a single noncommutative plane instead of two. The $2$ dimensional coadjoint orbits $^{c_{1}, c_{2}}\mathcal{O}^{0, 0, \tau}_{2}$ gave rise to the family of UIRs $U^{c_{1}, c_{2}}_{0, 0, \tau}$ as described in (\ref{taunonzero-rep-two-dim}). The corresponding Lie algebra representation on $L^{2}(\mathbb{R},dr)$ reads
    \begin{equation}\label{algbr-rep-tau-nonzero}
    \begin{split}
    &\hat{Q}_{1}=c_{1}\mathbb{I},\;\;
    \hat{Q}_{2}=c_{2}\mathbb{I},\\
    &\hat{P}_{1}=\kappa_{2}r,\;\;
    \hat{P}_{2}=-i\frac{\partial}{\partial r},
    \end{split}
    \end{equation}
    while the corresponding commutators are given by
    \begin{equation}\label{commutators-tau-nonzero}
    \begin{split}
    &[\hat{Q}_{1},\hat{P}_{1}]=[\hat{Q}_{2},\hat{P}_{2}]=0,\;\;
    [\hat{Q}_{1},\hat{Q}_{2}]=0,\\
    &[\hat{P}_{1},\hat{P}_{2}]=i\kappa_{2}\mathbb{I},\;\;
    [\hat{Q}_{1},\hat{P}_{2}]=[\hat{Q}_{2},P_{1}]=0.
    \end{split}
    \end{equation}
    Physically, (\ref{commutators-tau-nonzero}) refers to one of the two uncoupled noncommutative planes (see (\ref{algbr-rep-sigma-tau-nonzero})), the noncommutativity of which is measured by the dimensionless quantity $\kappa_{2}=-\tau\gamma$.

    \subsection{Case \texorpdfstring{$\rho= 0, \sigma\neq 0, \tau= 0.$}{sigma Nonzero}}\label{subsec:sigma-nonzero}
    The UIRs $U^{c_{3}, c_{4}}_{0, \sigma, 0}$, given by (\ref{sigmanonzero-rep-two-dim}), were found to be associated with the $2$ dimensional coadjoint orbits $^{c_{3}, c_{4}}\mathcal{O}^{0, \sigma, 0}_{2}$. We introduce the following operator of involution on $L^{2}(\mathbb{R},ds)$:
    \begin{equation}\label{invln-sigma-nonzero-rep}
    Tf(s)=f(-s),
    \end{equation}
    with $f\in L^{2}(\mathbb{R},ds)$.
    We then find a representation $\tilde{U}^{c_{3},c_{4}}_{0, \sigma, 0}$ unitarily equivalent to the one given by (\ref{sigmanonzero-rep-two-dim}), i.e. $T^{-1}U^{c_{3},c_{4}}_{0, \sigma, 0}T=\tilde{U}^{c_{3},c_{4}}_{0, \sigma, 0}$, with $\tilde{U}^{c_{3},c_{4}}_{0, \sigma, 0}$ given by
    \begin{eqnarray}\label{equiv-rep-sigmanonzero-two-dim}
    \lefteqn{(\tilde{U}^{c_{3}, c_{4}}_{0, \sigma, 0}(\theta,\phi,\psi,q_{1},q_{2},p_{1},p_{2})f)(s)}\nonumber\\
    &&=e^{ic_{3}q_{1}+ic_{4}q_{2}}e^{i\sigma(\phi+\beta p_{1}s-\frac{\beta}{2}p_{1}p_{2})}f(s-p_{2}),
    \end{eqnarray}
    where $f\in L^{2}(\mathbb{R}, ds)$.
    The corresponding representation of $\G$ on the same Hilbert space now reads
    \begin{equation}\label{algbr-rep-sigma-nonzero}
    \begin{split}
    &\hat{Q}_{1}=-\alpha_{1} s,\;\;
    \hat{Q}_{2}=i\frac{\partial}{\partial s},\\
    &\hat{P}_{1}=c_{3}\mathbb{I},\;\;
    \hat{P}_{2}=c_{4}\mathbb{I}.
    \end{split}
    \end{equation}
    The corresponding commutators read off immediately
    \begin{equation}\label{commutators-sigma-nonzero}
    \begin{split}
    &[\hat{Q}_{1},\hat{P}_{1}]=[\hat{Q}_{2},\hat{P}_{2}]=0,\;\;
    [\hat{Q}_{1},\hat{Q}_{2}]=i\kappa_{1}\mathbb{I},\\
    &[\hat{P}_{1},\hat{P}_{2}]=0,\;\;
    [\hat{Q}_{1},\hat{P}_{2}]=[\hat{Q}_{2},P_{1}]=0.
    \end{split}
    \end{equation}
    Physically, (\ref{commutators-sigma-nonzero}) corresponds to just one of the two uncoupled noncommutative planes (see (\ref{algbr-rep-sigma-tau-nonzero})) whose noncommutativity is controlled by the dimensionless parameter $\kappa_{1}=-\sigma\beta$.

    \subsection{Case \texorpdfstring{$\rho= 0, \sigma= 0, \tau= 0.$}{All zero}}\label{subsec:all-zero}
    The $0$ dimensional coadjoint orbits $^{c_{1}, c_{2}, c_{3}, c_{4}}\mathcal{O}^{0, 0, 0}_{0}$, admitting $1$ dimensional group representations given by (\ref{allzero-rep-zero-dim}), have the trivial algebra representation where all the basis elements of $\G$ are mapped to the scalar multiples of identity. The corresponding commutators are the same as those of the abelian group of translations in $\mathbb{R}^{4}$.

    This concludes the classification of all the families of unitary irreducible representations of $\G$ on appropriate Hilbert spaces. It is noteworthy that all possible representations of NCQM, as postulated in the multitude of examples found in the existing  physical literature (see, for example, \cite{delducetalt}), and the unitary irreducible representation of the Weyl-Heisenberg group for a quantum mechanical system of two degrees of freedom are all obtainable from the unitary dual of the {\em triply extended group of translations in $\mathbb{R}^{4}$}. However, this result is not surprising. The group $\g$ is a rather large group, compared to the Weyl-Heisenberg group. The Weyl-Heisenberg group only has one family of non-trivial unitary irreducible representations (one corresponding to each chosen value of the Planck constant). The group $\g$ has a vastly richer family of UIR's, and while the Weyl-Heisenberg group is not one of its subgroups, at the representation level its UIR's form a subset of those of $\g$. This fact also points up the universal nature of $\g$ as the underlying group of quantum mechanics. On the other hand, not all the representations of $\g$ necessarily have physical meanings or in particular, quantum mechanical significance. Consequently the generators in all these representations do not necessarily correspond to quantum mechanical position or momentum operators.

    \section{Various gauges of noncommutative quantum mechanics and their relation to \texorpdfstring{$\g$}{Triply extended group}}\label{sec:gauges-ncqm}

    Let us go back to the  representation $\tilde{U}^{\rho}_{\sigma,\tau}$ of
    $\g$, given in (\ref{unitary equivalence}), and the associated generators (\ref{algbr-rep-equivalent-nonzero}), obeying the commutation relations (\ref{commutators-rep-equivalent-nonzero}). As is well known (as shown for example in \cite{delducetalt}), there are other possible realizations of the operators $\hat{Q}_i , \hat{P}_i$, which also obey the same commutation relations, which can in many cases be related to the choice of a gauge in the following sense: the commutation relation, $[\hat{P}_1, \hat{P}_2] = i\bm{\mathcal{B}} \mathbb I=i\hbar B\mathbb{I}$, signals the presence of a constant magnetic field $B=\frac{\bm{\mathcal{B}}}{\hbar}$ in the system as has been discussed in section \ref{subsec:all-nonzero}. How this field, normal to $\hat{Q}_{1}\hat{Q}_{2}$-plane, can be obtained from the underlying vector potential, is given by (\ref{mag-vec-eq}). We already saw in section \ref{subsec:all-nonzero} that the {\em Landau gauge} corresponds to the choice  of vector potential given by $\vec{A}=(0,\frac{\bm{\mathcal{B}}}{\hbar}r_{1})=(0,Br_{1})$. It can be clearly seen that one could choose $\vec{A}$ as $\vec{A}=(-\frac{B}{2}r_{2},\frac{B}{2}r_{1})$ and still satisfy (\ref{mag-vec-eq}). This gauge is known as the {\em symmetric gauge} in the physical literature. A change of gauge for the vector potential, in general, does not affect the physics of the system. At the quantum mechanical level a change of gauge affects the exact realization of the  operators $\hat{Q}_i , \hat{P}_i$, without altering the commutation relation $[\hat{P}_1, \hat{P}_2] = i\bm{\mathcal{B}} \mathbb I$. Furthermore, the differently realized generators would then lift up to unitarily equivalent representations of $\g$. As mentioned in section \ref{subsec:all-nonzero}, the realization given in (\ref{algbr-rep-equivalent-nonzero}) corresponds to the Landau gauge. Now we look at the so-called symmetric gauge of the underlying vector potential from a representation theoretic point of view.

    We  have the following theorem
    \begin{Theo}\label{sym-gauge-thm}
    The nilpotent Lie group $\g$, obeying the group law (\ref{grp-law}), admits a unitary irreducible representation $\mathcal{U}_{\hbox{\tiny{sym}}}$ given by
    \begin{eqnarray}\label{symmetric-gauge-rep}
    \lefteqn{(\mathcal{U}_{\hbox{\tiny{sym}}}(\theta,\phi,\psi,q_{1},q_{2},p_{1},p_{2})f)(r_{1},r_{2})}\nonumber\\
    &&=e^{i(\theta+\phi+\psi)}e^{i\left[\alpha p_{1}r_{1}+\alpha p_{2}r_{2}-\frac{\alpha(\alpha-\sqrt{\alpha^{2}-\beta\gamma})}{\beta}(q_{1}r_{2}-q_{2}r_{1})+\frac{\sqrt{\alpha^{2}-\beta\gamma}}{2}(p_{1}q_{1}+p_{2}q_{2})\right]}\nonumber\\
    &&\times f\left(r_{1}-\frac{\beta}{2\alpha}p_{2}+\frac{\alpha+\sqrt{\alpha^{2}-\beta\gamma}}{2\alpha}q_{1},r_{2}+\frac{\beta}{2\alpha}p_{1}+\frac{\alpha+\sqrt{\alpha^{2}-\beta\gamma}}{2\alpha}q_{2}\right),\label{sym-rep-exprssn}
    \end{eqnarray}
    with $f\in L^{2}(\mathbb{R}^{2}, dr_{1}dr_{2})$. This representation is unitarily equivalent to $\tilde{U}^{\rho}_{\sigma,\tau}$ given by (\ref{nonzero-equivalent-rep}) for $\rho=\sigma=\tau=1$.
    \end{Theo}

The proof is given in the Appendix.

    We choose $\alpha=\frac{1}{\hbar}$ in (\ref{symmetric-gauge-rep}). One can verify that this choice is dimensionally consistent by looking at (\ref{grp-law}) or (\ref{symmetric-gauge-rep}). Hence, we take $C=\frac{1}{\alpha}$ in (\ref{generator-formula}) and obtain the corresponding unitary irreducible representation of the noncentral elements of $\G$ on $L^{2}(\mathbb{R}^{2}, dr_{1}dr_{2})$ which is as follows
    \begin{equation}\label{sym-gauge-algbr-rep}
    \begin{split}
    &\hat{Q}_{1}=r_{1}+\frac{i\vartheta}{2}\frac{\partial}{\partial r_{2}},\\[4pt]
    &\hat{Q}_{2}=r_{2}-\frac{i\vartheta}{2}\frac{\partial}{\partial r_{1}},\\[4pt]
    &\hat{P}_{1}=\frac{(\hbar-\sqrt{\hbar^{2}-\bm{\mathcal{B}}\vartheta})}{\vartheta}r_{2}-\frac{i(\hbar+\sqrt{\hbar^{2}-\bm{\mathcal{B}}\vartheta})}{2}\frac{\partial}{\partial r_{1}},\\[4pt]
    &\hat{P}_{2}=\frac{(\sqrt{\hbar^{2}-\bm{\mathcal{B}}\vartheta}-\hbar)}{\vartheta}r_{1}-\frac{i(\hbar+\sqrt{\hbar^{2}-\bm{\mathcal{B}}\vartheta})}{2}\frac{\partial}{\partial r_{2}},
    \end{split}
    \end{equation}
    and the central elements of the algebra are all mapped to the scalar multiple of the identity of the underlying Hilbert space. Here, in addition to taking $\alpha=\frac{1}{\hbar}$, we have chosen $\beta=-\frac{\vartheta}{\hbar^{2}}$ and $\gamma=-\frac{\bm{\mathcal{B}}}{\hbar^{2}}$. The representation (\ref{sym-gauge-algbr-rep}) is easily seen to satisfy the set of commutation relations given by (\ref{commutators-rep-equivalent-nonzero}). As indicated earlier, this representation is due to the choice of the {\em symmetric gauge} (see \cite{delducetalt} for details) for the underlying vector potential.

    Also, we assume that $\rho^{2}\alpha^{2}-\gamma\beta\sigma\tau\neq 0$ holds in course of deriving (\ref{nonzero-equivalent-rep}). And since $\rho=\sigma=\tau=1$, we have $\alpha^{2}-\beta\gamma\neq 0$. Equation (\ref{sym-gauge-algbr-rep}) is compatible with this inequality. It is also important to note that in the limiting situation where $\alpha^{2}-\beta\gamma\rightarrow 0$ or equivalently $\hbar^{2}-\bm{\mathcal{B}}\theta\rightarrow 0$, (\ref{sym-gauge-algbr-rep}) becomes reducible as $\hat{P}_{2}+\frac{\hbar}{\vartheta}\hat{Q}_{1}\rightarrow 0$ and $\hat{P}_{1}-\frac{\hbar}{\vartheta}\hat{Q}_{2}\rightarrow 0$. In other words, $\hat{Q}_{1}$ and $\hat{Q}_{2}$ become proportional to $\hat{P}_{2}$ and $\hat{P}_{1}$, respectively, in this limiting situation.

    Now, in a similar manner, other unitarily equivalent realizations of the commutation relations (\ref{commutators-rep-equivalent-nonzero}) may be obtained by using other gauge equivalent  vector potentials. It is also clear from (\ref{algbr-rep-equivalent-nonzero}) and (\ref{sym-gauge-algbr-rep}) that the two sets of operators $\hat{Q}_{i}, \hat{P}_{i}, \; i=1,2,$ appearing in those two sets of equations are related by a linear transformation. It is therefore natural to ask what is the largest set of such transformations which would leave the commutation relations (\ref{commutators-rep-equivalent-nonzero}) invariant. This question is answered in the following section.

    \section{Group of transformations preserving the commutation relations of noncommutative quantum mechanics}\label{sec:grp-trans-prsrv}
 It is a well-known fact that in classical mechanics the set of transformations which preserve the canonical Poisson brackets between the phase space variables $p_i$ and $q_j$ in $\mathbb{R}^{2n}$, form the Lie group $Sp(2n,\mathbb{R})$. In standard quantum mechanics the canonical commutation relations are also invariant under this same group. For the noncommutative system of two degrees of freedom, the phase space is $\mathbb{R}^{4}$ and the transformations between two different sets of  $\{\hat{Q}_i , \hat{P}_i\}, \; i=1,2$, obeying the same commutation relations (\ref{commutators-rep-equivalent-nonzero}), also form a group that is  isomorphic to $Sp(4,\mathbb{R})$, as will follow from the following  considerations.

    Consider two sets of phase space variables ${\hat{Q}_{1}, \hat{P}_{2}, \hat{Q}_{2}, \hat{P}_{1}}$ and ${\hat{Q}^{\prime}_{1}, \hat{P}^{\prime}_{2}, \hat{Q}^{\prime}_{2}, \hat{P}^{\prime}_{1}}$ in $\mathbb{R}^{4}$ satisfying the commutation relations (\ref{commutators-rep-equivalent-nonzero}). Let $\mathbb{M}$ be a $4\times 4$ matrix, with real entries, for which
    \begin{equation}\label{trnsfrmtn-matrx-def}
    \begin{bmatrix}\hat{Q}^{\prime}_{1}\\ \hat{P}^{\prime}_{2}\\ \hat{Q}^{\prime}_{2}\\ \hat{P}^{\prime}_{1}\end{bmatrix}=\mathbb{M}\begin{bmatrix}\hat{Q}_{1}\\ \hat{P}_{2}\\ \hat{Q}_{2}\\ \hat{P}_{1}\end{bmatrix}
    \end{equation}
    We then have the following theorem:
    \begin{Theo}\label{thm-trnsfrm-matrix}
    The $4\times 4$ real matrices $\mathbb{M}$ in (\ref{trnsfrmtn-matrx-def}), preserving the commutation relations (\ref{commutators-rep-equivalent-nonzero}) of a general non-commutative quantum system of two degrees of freedom, satisfy the  condition
    \begin{equation}\label{trnsfrmn-mtrx-formula}
    \mathbb{M}\mathbb{Q}\mathbb{M}^{T}=\mathbb{Q},
    \end{equation}
    where  $\mathbb{Q}$ is the $4\times $ block off-diagonal matrix,
    \begin{equation}\label{block-form}
    \mathbb{Q}=\begin{bmatrix}0&Q\\-Q^{T}&0\end{bmatrix},
    \end{equation}
    with $2\times 2$ matrix $Q$ given by the $2\times 2$ matrix
    \begin{equation}\label{two-by-two-matrx}
    Q=\begin{bmatrix}-\frac{\vartheta}{\hbar}&-1\\1&\frac{\bm{\mathcal{B}}}{\hbar}\end{bmatrix}.
    \end{equation}
    \end{Theo}

    The proof is given in the Appendix.

    \begin{remark}
    A few remarks are in order. The converse of Theorem (\ref{thm-trnsfrm-matrix}) is also true. As a result, (\ref{trnsfrmn-mtrx-formula}) is a necessary and sufficient condition for the noncommutative commutation relations to be preserved. Also, the $2\times 2$ matrix $Q$, given by (\ref{two-by-two-matrx}), is required to be invertible, i.e. $\hbar^{2}-\bm{\mathcal{B}}\vartheta\neq 0$, a fact that has also been exploited in \cite{delducetalt}. Finally, all $4\times 4$ real matrices $\mathbb M$ , satisfying (\ref{trnsfrmn-mtrx-formula}), can easily be verified to form a group under matrix multiplication. Actually, as shown below, these matrices form a real Lie group, hence forth denoted by $\mathfrak{S}(4,\mathbb{R})$.
    \end{remark}

    We have the following result concerning group isomorphism.

    \smallskip

    \textbf{Result:}
     The $10$ dimensional real Lie group $\mathfrak{S}(4,\mathbb{R})$ is isomorphic to the simple Lie group $Sp(4,\mathbb{R})$. The isomorphism $f:\mathfrak{S}(4,\mathbb{R})\rightarrow Sp(4,\mathbb{R})$, can be written as $f(\mathbb{M})=\mathcal{U}^{-1}\mathbb{M}\mathcal{U}$, where $\mathcal U$ is the  $4\times 4$ invertible matrix:
    \begin{equation}\label{Invertble-matrx}
    \mathcal{U}=\begin{bmatrix}-1&\frac{\vartheta}{\hbar}&0&0\\\frac{\bm{\mathcal{B}}}{\hbar}&-1&0&0\\0&0&1&0\\0&0&0&1\end{bmatrix}.
    \end{equation}

The isomorphism $f$ in this context is what one expects to follow naturally because the relevant operators representing the noncentral generators of $\g$ can be expressed as linear combinations of those which generate the CCR of standard quantum mechanics on $L^{2}(\mathbb{R}^{2}, dr_{1}dr_{2})$. Also, $Sp(4,\mathbb{R})$ is the group of transformations that preserve the CCR for a system with $2$ degrees of freedom. The $4 \times 4$ matrix $\mathcal{U}$ in (\ref{Invertble-matrx}) is actually the matrix of transformation between the standard quantum mechanical and noncommutative quantum mechanical (in this case, Landau gauge) representations as can be readily seen from (\ref{algbr-rep-equivalent-nonzero}). We could also have chosen $\mathcal{U}$ as the one arising from the symmetric gauge (\ref{sym-gauge-algbr-rep}). Thus, the choice of $\mathcal{U}$ is evidently not unique.

\begin{remark}
That the matrix group $\mathfrak{S}(4,\mathbb{R}$) will be isomorphic to $Sp(4,\mathbb{R})$, is a simple consequence of Linear Darboux theorem which guarantees the existence of a suitable Darboux basis that will transform any symplectic form in $\mathbb{R}^{4}$ into the standard one.
\end{remark}

As a concrete example of $\mathbb{M}$, introduced in (\ref{trnsfrmtn-matrx-def}), let us consider the phase space variables associated with the Landau gauge (see \ref{algbr-rep-equivalent-nonzero}) and symmetric gauge (see \ref{sym-gauge-algbr-rep})  with
    \begin{equation}\label{example-matrix}
    \begin{bmatrix}\hat{Q}_{1}\\ \hat{P}_{2}\\ \hat{Q}_{2}\\ \hat{P}_{1}\end{bmatrix}^{\hbox{\tiny{sym}}}=\mathbb{M}\begin{bmatrix}\hat{Q}_{1}\\ \hat{P}_{2}\\ \hat{Q}_{2}\\ \hat{P}_{1}\end{bmatrix}^{\hbox{\tiny{Landau}}}
    \end{equation}.

After some straightforward but rather lengthy computations, one arrives at
\begin{equation}\label{explicit-mtrx-frm}
\mathbb{M}=\begin{bmatrix}1+\frac{\bm{\mathcal{B}}\vartheta}{2(\hbar^{2}-\bm{\mathcal{B}}\vartheta)}&\frac{\vartheta\hbar}{2(\hbar^{2}-\bm{\mathcal{B}}\vartheta)}&0&0\\\frac{\bm{\mathcal{B}}(2\hbar\sqrt{\hbar^{2}-\bm{\mathcal{B}}\vartheta}-\bm{\mathcal{B}}\vartheta)}{2(\hbar^{2}-\bm{\mathcal{B}}\vartheta)(\hbar+\sqrt{\hbar^{2}-\bm{\mathcal{B}}\vartheta})}&\frac{\hbar(3\sqrt{\hbar^{2}-\bm{\mathcal{B}}\vartheta}-\hbar)}{2(\hbar^{2}-\bm{\mathcal{B}}\vartheta)}&0&0\\0&0&1&\frac{\vartheta}{2\hbar}\\0&0&\frac{\hbar-\sqrt{\hbar^{2}-\bm{\mathcal{B}}\vartheta}}{\vartheta}&\frac{\hbar+\sqrt{\hbar^{2}-\bm{\mathcal{B}}\vartheta}}{2\hbar}\end{bmatrix}.
\end{equation}
    It can, then, be immediately verified that $\mathbb{M}$, given by (\ref{explicit-mtrx-frm}), indeed satisfies (\ref{trnsfrmn-mtrx-formula}).

 \section{Relationship with complex Hermite polynomials}\label{Sec:Hermt-pol}
 We explore in this section a connection between a model of non-commutative quantum mechanics, governed by a certain restricted version of the commutation relations (\ref{commutators-rep-equivalent-nonzero}), and a family of {\em deformed complex Hermite polynomials\/.} We note first of all, that an irreducible representation of the commutation relations
 \begin{equation}\label{algbr-crtn-ann-indp}
    [a_i,a_j]=[a_{i}^{\dag},a_{j}^{\dag}]=0,\;\; [a_{i},a_{j}^{\dag}]=\delta_{ij}\mathbb{I},\;\;i,j=1,2,
    \end{equation}
of a standard quantum mechanical system for two degrees of freedom, can be constructed on the Hilbert space $L^2 (\mathbb C , e^{-\vert z\vert^2}\; \frac {dx\;dy}{\pi} )$ as
\be
 a_1 = \partial_z ,\qquad a_1^\dag = z - \partial_{\overline{z}}, \qquad\qquad
 a_2 = \partial_{\overline{z}}, \qquad a_2^\dag = \overline{z} - \partial_z \; .
\label{ord-crann-comp}
\en

We denote by $\mathbf I$ the constant function in $L^2 (\mathbb C ,
e^{-\vert z\vert^2}\; \frac {dx\;dy}{\pi} )$, which is equal to one everywhere.
Then the vectors,
\be
  H_{n,k} = \frac {(a_1^\dag )^n\; (a_2^\dag)^k}{\sqrt{n!\; k!}}\mathbf I\; , \qquad
  n,k = 0,1,2, \ldots , \infty
\label{hkl-vects}
\en
form an orthonormal basis of $L^2 (\mathbb C , e^{-\vert z\vert^2}\; \frac {dx\;dy}{\pi} )$.
It can be shown that
\be
 H_{n,k} (z, \overline{z})  = \frac{(-1)^{n+k}}{\sqrt{n!\;k!}} \;e^{\vert z\vert^2}\partial^n_z \partial^k_{\overline z}
   \; e^{-\vert z\vert^2}\; ,
\label{comp-herm-poly}
\en
or explicitly,
\be
  H_{n,k} (z, \overline{z}) = \sqrt{n!\;k!} \sum_{j= 0}^{n\curlyvee k}\frac {(-1)^j}{j!}
    \frac {(\overline{z})^{n-j}}{(n-j)!}\; \frac {z^{k-j}}{(k-j)!}\; ,
\label{comp-herm-poly9}
\en
where $n\curlyvee k$ denotes the smaller one of the two integers $n$ and $k$.
The functions $H_{n,k} (z, \overline{z})$ are known in the literature (see, for example \cite{ghanmi,intissar,Ito,Shigekawa}) as the  {\em complex Hermite polynomials}. They form a basis in
$L^2 (\mathbb C , e^{-\vert z\vert^2}\; \frac {dx\;dy}{\pi} )$ and satisfy the orthonormality condition
\be
  \int_{\mathbb C}\overline{H_{n,k} (z, \overline{z})}H_{m,l} (z, \overline{z})\;e^{-\vert z\vert^2}\;
     \frac {dx\;dy}{\pi} = \delta_{nm}\;\delta_{kl}\; .
\label{comp-herm-orthog}
\en
From the way we have introduced them here, it is clear that these polynomials are the ones naturally associated to a standard quantum mechanical system of two degrees of freedom (or with two independent oscillators).

  Consider now a non-commutative quantum system obeying the commutation relations (\ref{commutators-rep-equivalent-nonzero}) and let us assume that we are in the symmetric gauge (\ref{sym-gauge-algbr-rep}). We define the {{\em deformed creation and annihilation operators\/,} using the operators $\hat{Q}_i, \; \hat{P}_i, \;\; i = 1,2,$ in (\ref{sym-gauge-algbr-rep}),
\bea
  a_{i}^{\mbox{\small{nc}\dag}}& = &\sqrt{\frac{M\Omega}{2\hbar}}\left(\hat{Q}_i-\frac i{M\Omega}\hat{P}_i \right),\nonumber\\
  a_{i}^{\mbox{\small{nc}}}& = &\sqrt{\frac{M\Omega}{2\hbar}}\left(\hat{Q}_i +\frac i{M\Omega}\hat{P}_i \right)\; ,\quad  i =1,2\; ,
\label{nc-cr-ann}
\ena
where the $M$ and $\Omega$ are a mass and an angular frequency parameter, which can be adjusted later. These operators are seen obey the commutation relations
\begin{equation}\label{commu-rel-NCHO}
    \begin{split}
     [a_{i}^{\mbox{\small{nc}}},a_{j}^{\mbox{\small{nc}\dag}}] &= \delta_{ij}\mathbb{I}+\frac{i\epsilon_{ij}}{2\hbar}M\Omega\left(\vartheta+
     \frac{\bm{\mathcal{B}}}{M^{2}\Omega^{2}}\right)\mathbb{I}\\
[a_{i}^{\mbox{\small{nc}}},a_{j}^{\mbox{\small{nc}}}] &= \frac{i\epsilon_{ij}}{2\hbar}M\Omega\left(\vartheta-\frac{\bm{\mathcal{B}}}{M^{2}
\Omega^{2}}\right)\mathbb{I},
    \end{split}
    \end{equation}
    where $i,j=1,2$ and $\epsilon_{ij}$ is the totally antisymmetric symbol.
Since we want the two operators $a_{i}^{\mbox{\small{nc}}}, \; i =1,2,$ to still represent independent bosons, we impose the condition that the second commutator above be zero. This implies taking
\begin{equation}\label{constrnt-eqn}
    \vartheta=\frac{\bm{\mathcal{B}}}{M^{2}\Omega^{2}}.
    \end{equation}
    and hence the other commutator now reads
    \begin{equation}\label{rest-eq}
    [a_{i}^{\mbox{\small{nc}}},a_{j}^{\mbox{\small{nc}\dag}}] = \delta_{ij}\mathbb{I}+\frac{i\epsilon_{ij}\vartheta M\Omega}{\hbar}\mathbb{I},
    \end{equation}
which still means that we are in the framework of noncommutative quantum mechanics, since $[a_{1}^{\mbox{\small{nc}}},a_{2}^{\mbox{\small{nc}\dag}}] \neq 0$.

 We next introduce the standard creation and annihilation operators (obeying the commutation relations (\ref{algbr-crtn-ann-indp})), in terms of the usual position and momentum operators of quantum mechanics,
\bea
  a_{i}^\dag& = &\sqrt{\frac{m\omega}{2\hbar}}\left(Q_i-\frac i{m\omega}P_i \right)\nonumber\\
  a_{i}& = &\sqrt{\frac{m\omega}{2\hbar}}\left(Q_i +\frac i{m\omega}P_i \right)\;, \quad i=1,2,
\label{st-cr-ann}
\ena
where
\be
  m\omega = \frac {2\hbar M\Omega}{\hbar + \sqrt{\hbar^{2} - \bm{\mathcal{B}} \vartheta}} =
  \frac {2\hbar\sqrt{\bm{\mathcal{B}}}}{\sqrt{\vartheta}(\hbar + \sqrt{\hbar^{2} - \bm{\mathcal{B}} \vartheta})}\; .
\label{osc-mass}
\en
A straightforward computation, using (\ref{sym-gauge-algbr-rep}) then gives
\bea
  a_{1}^{\mbox{\small{nc}\dag}}& = & \sqrt{\nu} a_1^{\dag} -i\sqrt{1-\nu}a_2^{\dag} \nonumber\\
  a_{2}^{\mbox{\small{nc}\dag}}& = &i\sqrt{1-\nu}a_1^{\dag} + \sqrt{\nu}a_2^{\dag}, \qquad
  \nu = \frac {\hbar + \sqrt{{\hbar}^{2} - \bm{\mathcal{B}} \vartheta}}{2\hbar}\; .
\label{nc-cr-ann2}
\ena

Note that in (\ref{nc-cr-ann2}), as $\hbar^{2}-\bm{\mathcal{B}}\vartheta\rightarrow 0$, $a_{1}^{\mbox{\small{nc}\dag}}$ and $a_{2}^{\mbox{\small{nc}\dag}}$ are no longer linearly independent of each other and hence the representation becomes reducible in this limiting case. This reducibility of the limiting situation of the deformed annihilation operators agrees with the discussion given in section \ref{sec:gauges-ncqm}. The agreement here is what one expects to follow naturally since the symmetric gauge representation (\ref{symmetric-gauge-rep}) is unitarily equivalent to the one given by (\ref{nonzero-equivalent-rep}) for $\rho=\sigma=\tau=1$. And one uses the fact that $\rho^{2}\alpha^{2}-\gamma\beta\sigma\tau\neq 0$ holds while deriving the unitary irreducible representations (\ref{nonzero-equivalent-rep}) of $\g$.

It is now interesting and useful to realize the operators (\ref{nc-cr-ann2}) using the complex representation (\ref{ord-crann-comp}) and to look at the {\em deformed complex polynomials}
\be
  H_{n ,k}^{\mbox{\small{nc}}} = \frac {(a_{1}^{\mbox{\small{nc}}\dag})^n\;(a_{2}^{\mbox{\small{nc}}\dag})^k}{\sqrt{n!\; k!}}\mathbf I\; , \qquad
  n,k = 0,1,2, \ldots , \infty\; ,
\label{nc-hkl-vects}
\en
in analogy with (\ref{hkl-vects}). These polynomials do not satisfy an orthogonality relation of the type (\ref{comp-herm-orthog}). However, it has been shown in \cite{alibaloghshah} that there exists a dual set of polynomials $\widetilde{H}_{n ,k}^{\mbox{\small{nc}}}$ for which one has the {\em biorthogonality relation}
\be
  \int_{\mathbb C}\overline{\widetilde{H}^{\mbox{\small{nc}}}_{n,k} (z, \overline{z})}H^{\mbox{\small{nc}}}_{m,l} (z, \overline{z})\;e^{-\vert z\vert^2}\;
     \frac {dx\;dy}{\pi} = \delta_{nm}\;\delta_{kl}\; .
\label{comp-herm-biorthog}
\en

We can go further  and define deformed creation and annihilation operators using an arbitrary $GL(2, \mathbb C)$ matrix
$$ g = \begin{pmatrix} g_{11} & g_{12} \\ g_{21} & g_{22} \end{pmatrix} $$
in the manner
    \begin{equation}\label{defrmd-crtn-annhltn-oprtr}
    a_{1}^{g\dag}=g_{11}a_{1}^{\dag}+g_{21}a_{2}^{\dag},\;\;\;\;\;a_{2}^{g\dag}=g_{12}a_{1}^{\dag}+g_{22}a_{2}^{\dag}.
    \end{equation}
and construct the corresponding deformed polynomials $H^g_{n,k}$. In this case the dual polynomials are obtained using the matrix $\widetilde{g} = (g^\dag)^{-1}$. We emphasize here that the term deformation is used here in the specific sense that we are using the deformed ladder operators (\ref{defrmd-crtn-annhltn-oprtr}), which are linear combinations of the ladder operators of standard quantum mechanics. The term deformed polynomials are consequently the ones constructed using these deformed ladder operators.

To relate these more general polynomials to noncommutative quantum mechanics one still has to impose the condition $[a^g_1 , a^g_2] = 0$. Thus, generally, one gets a matrix of the form
\begin{equation}\label{eq-g-mtrx}
g=\begin{bmatrix}re^{i\kappa}&\sqrt{1-r^{2}}e^{i\{\kappa+\epsilon(r)\}}\\\sqrt{1-r^{2}}e^{i\delta}&-re^{i\{\delta-\epsilon(r)\}}\end{bmatrix}.
    \end{equation}
    Here, $\epsilon(r)$, being a function of $r$, is given by

    \begin{equation}\label{epsilon-func}
    \epsilon(r)=\arcsin\left(\frac{\vartheta M\Omega}{2\hbar r\sqrt{1-r^{2}}}\right).
    \end{equation}

    From (\ref{eq-g-mtrx}), one finds that  $0<r\leq 1$. But the condition $-1\leq\frac{\vartheta M\Omega}{2\hbar r\sqrt{1-r^{2}}}\leq 1$ puts further restrictions on $r$, requiring that
    \begin{equation}\label{r-bound-eq} r\in\left[\sqrt{\frac{1}{2}-\sqrt{\frac{1}{4}-\frac{\vartheta^{2}M^{2}
    \Omega^{2}}{4\hbar^{2}}}},\sqrt{\frac{1}{2}+\sqrt{\frac{1}{4}-
    \frac{\vartheta^{2}M^{2}\Omega^{2}}{4\hbar^{2}}}}\right],
    \end{equation}
along with, $0<\frac{\vartheta M\Omega}{\hbar}\leq 1$. Also, $\kappa\in[-\epsilon(r),2\pi-\epsilon(r))$ and $\delta\in[\epsilon(r),2\pi+\epsilon(r))$, where, $\epsilon(r)\in\left[\arcsin\left(\frac{\vartheta M\Omega}{\hbar}\right),\frac{\pi}{2}\right]$, as a transcendental function of $r$, varies according to (\ref{epsilon-func}).

    To summarize, if we consider the operators $\hat{Q}_1$, $\hat{Q}_2$, $\hat{P}_1$ and $\hat{P}_2$, in the {\em symmetric gauge representation} (\ref{sym-gauge-algbr-rep}) of the {\em triply extended algebra of translations} $\G$, to be the respective positions and momenta of the two bosons of the underlying coupled system and impose a constraint given by (\ref{constrnt-eqn}), then the resulting creation and annihilation operators are  linear combinations of the canonical creation and annihilation operators via the invertible matrix
\begin{equation}\label{restricted-symgauge}
g_{\hbox{\tiny{sym}}}=\begin{bmatrix}\sqrt{\nu}&i\sqrt{1-\nu}\\-i\sqrt{1-\nu}&\sqrt{\nu}\end{bmatrix}.
\end{equation}
It is easy to see that $g_{\hbox{\tiny{sym}}}$, given by (\ref{restricted-symgauge}), is a special case of the matrix $g$ introduced in (\ref{eq-g-mtrx}) with $r^{2}=\nu$, $\kappa=0$, $\delta=\frac{3\pi}{2}$, and $\epsilon=\frac{\pi}{2}$.

Before closing this section, we  examine the geometric consequences of (\ref{constrnt-eqn}) from the representation theoretic point of view. To this end, (\ref{constrnt-eqn}) together with (\ref{idntfctn-nonzero-rep}) imply
    \begin{equation}\label{geom-consqnc}
    \tau=\frac{\beta M^{2}\Omega^{2}}{\gamma}\sigma:=K_{g}\sigma.
    \end{equation}
Also, together with (\ref{idntfctn-nonzero-rep}), the inequality $0<\frac{\vartheta M\Omega}{\hbar}\leq 1$, as has already been mentioned earlier in this section, puts severe restrictions both on $\sigma$ and $\rho$:
\begin{equation}\label{geom-consqnc-contnd}
\sigma\in(-\infty,\infty)\;\;,\;\;\rho\geq-\sigma\left(\frac{\beta M\Omega}{\alpha}\right)\;\;\mbox{and}\;\;-\frac{\sigma}{\rho}\left(\frac{\beta M\Omega}{\alpha}\right)>0.
\end{equation}
Considering the fact that $\alpha$, $\beta$ and $M\Omega$ are all positive constants, the last condition, in (\ref{geom-consqnc-contnd}), simply states that $\sigma$ and $\rho$ have to be of opposite sign. The equality in the second relation of (\ref{geom-consqnc-contnd}) insinuates the pertinence of a family of $2$-dimensional coadjoint orbits $^{\kappa,\delta}\mathcal{O}^{\rho,\zeta_{\hbox{\scalebox{.45}{HO}}}}_{2}$ with $\zeta_{\hbox{\scalebox{.45}{HO}}}=-\frac{\beta M\Omega}{\alpha}$ (see page 7, section \ref{sec:coad-orbts} for the notation) in the context of the coupled bosonic system. The strict inequality, on the other hand, in the second relation of (\ref{geom-consqnc-contnd}) suggests that there is an associated family of $4$-dimensional coadjoint orbits $\mathcal{O}^{\rho,\sigma,K_{g}\sigma}_{4}$ with $\rho$ and $\sigma$ satisfying the inequality in question and $K_{g}$ is as in (\ref{geom-consqnc}). As a whole, equation (\ref{geom-consqnc-contnd}) can be geometrically interpreted  using a diagram given by figure \ref{fig:figsecond}.

\begin{figure}[thb]
\centering
\includegraphics[width=8cm]{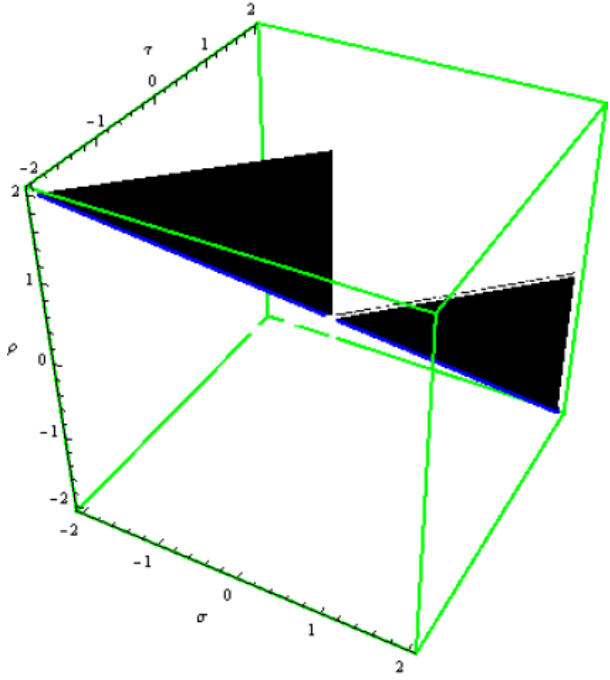}
\caption{The surface in $\mathbb{R}^{3}_{0}$ that relates to the representations associated with the $g$-deformed complex Hermite polynomials. Here, $\alpha$, $\beta$, $\gamma$ and $M\Omega$ are all taken to be 1 in their appropriate units.}
\label{fig:figsecond}
\end{figure}

To sum up, a family of $2$-dimensional coadjoint orbits $^{\kappa,\delta}\mathcal{O}^{\rho,\zeta_{\hbox{\scalebox{.45}{HO}}}}_{2}$ and $4$-dimensional coadjoint orbits $\mathcal{O}^{\rho,\sigma,K_{g}\sigma}_{4}$ along with the associated unitary irreducible representations of the {\em triply extended group of translations} $\g$ (see \ref{nonzero-equivalent-rep}) describe the coupled bosonic system under study. How the $2$ and $4$-dimensional coadjoint orbits relate to an open subset $\mathbb{R}^{3}_{0}\subset\mathbb{R}^{3}$ are all discussed in section \ref{sec:coad-orbts}. In figure \ref{fig:figsecond}, the $2$-dimensional subset of $\mathbb{R}^{3}_{0}$, colored black, corresponds to the $4$-dimensional coadjoint orbits $\mathcal{O}^{\rho,\sigma,K_{g}\sigma}_{4}$ satisfying (\ref{geom-consqnc-contnd}) with the second relation being a strict inequality. And, the associated unitary irreducible representations are the ones that contribute to the coupled bosonic system under study. The $2$ blue half-lines in figure \ref{fig:figsecond}, on the other hand, give rise to $2$-dimensional coadjoint orbits $^{\kappa,\delta}\mathcal{O}^{\rho,\zeta_{\hbox{\scalebox{.45}{HO}}}}_{2}$ corresponding to the coupled bosonic system. It is illustrated in figure \ref{fig:figthird} that the curved surface in figure \ref{fig:figfirst}, indeed intersects the flat surface of figure \ref{fig:figsecond} along the $2$ half lines of figure \ref{fig:figsecond}.

\begin{figure}[thb]
\centering
\includegraphics[width=8cm]{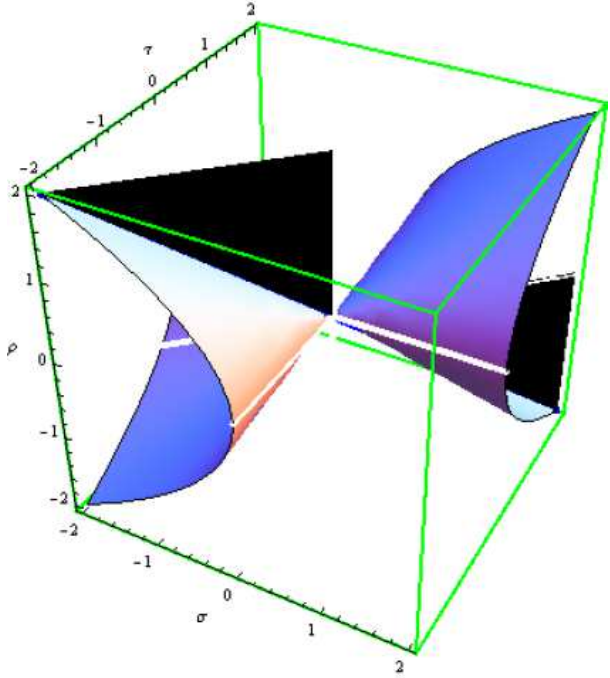}
\caption{In $\mathbb{R}^{3}_{0}$, the curved surface, corresponding to the 2-dimensional coadjoint orbits, intersects the flat surface associated with the coupled bosonic system along 2 disjoint half lines.}
\label{fig:figthird}
\end{figure}

It is important to note that $K_{g}=\frac{\beta M^{2}\Omega^{2}}{\gamma}$, in (\ref{geom-consqnc}), is a dimensionless coefficient. It is taken to be $1$ in figure \ref{fig:figsecond}. Also, $\rho$ and $\sigma$ can take values on the real line in accordance with (\ref{geom-consqnc-contnd}). In a physically meaningful setting, using (\ref{idntfctn-nonzero-rep}), one observes that $\rho\in(0,\infty)$ and each of $\sigma$ and $\tau$ has to be strictly negative. In this case, we just have the solid triangle for $\rho>0$ in figure \ref{fig:figsecond} with one blue half line as one of the edges.

    \section{Conclusion and future perspectives}
    In this paper, we have shown that the {\em triply extended group of translations in $\mathbb{R}^{4}$}, $\g$,  contains various representations, associated with different gauges of noncommutative quantum mechanics (see \cite{delducetalt}), viz., the Landau  and symmetric gauges, in its unitary dual. The unitary irreducible representations of standard quantum mechanics are also sitting inside its unitary dual. The representations associated with a coupled bosonic system, that give rise to certain deformed complex Hermite polynomials (see (\cite{aliismailshah}) and (\cite{alibaloghshah}) for detail), are just a family of unitary irreducible representations of $\g$. The relevant coadjoint orbits of $\g$, sitting inside the $7$ dimensional dual Lie algebra, have all been identified.
    The second cohomology group of the group of translations in $\mathbb{R}^4$ is a $6$ dimensional vector space. In this paper, we considered $[\hat{Q}_{1},\hat{P}_{1}]$, $[\hat{Q}_{2},\hat{P}_{2}]$, $[\hat{Q}_{1},\hat{Q}_{2}]$, and $[\hat{P}_{1},\hat{P}_{2}]$ to be nonvanishing in general, as is done in NCQM. The strengths of the first two noncommutativity were chosen to be the same in order to preserve the structure of standard quantum mechanics. Along with this quantum mechanical noncommutativity, the position and momentum noncommutativity give rise to three independent central extensions of the abelian group of translations in $\mathbb{R}^4$. While the goal of the present paper was to study the role of this triply extended group $\g$ in NCQM, it would be interesting to study the other extensions of the group of translations in physically meaningful contexts, e.g. rotational invariance. In the present paper, we restricted ourselves to $2$ degrees of freedom meaning that we studied $2$ dimensional NCQM from a group-theoretic point of view. But one could study possible extensions of the theory to quantum systems with additional degrees of freedom as well and look for a more general theory by constructing a more general version of $\g$.

    \section{Appendix}\label{sec-app}
In this Appendix we collect together the proofs of some of the results quoted in the paper.
\medskip

\prf {\bf of Theorem \ref{sym-gauge-thm}}

By a rather straightforward but fairly lengthy computation, one can show that $\mathcal{U}_{\hbox{\tiny{sym}}}$, given by (\ref{sym-rep-exprssn}), is indeed a representation of $\g$. In other words, the following holds
\begin{align}\label{rep-proof-fin}
    &(\mathcal{U}_{\hbox{\tiny{sym}}}((\theta,\phi,\psi,q_{1},q_{2},p_{1},p_{2})(\theta^{\prime},\phi^{\prime},\psi^{\prime},q_{1}^{\prime},q_{2}^{\prime},p_{1}^{\prime},p_{2}^{\prime}))f)(r_{1},r_{2})\nonumber\\
    &=(\mathcal{U}_{\hbox{\tiny{sym}}}(\theta,\phi,\psi,q_{1},q_{2},p_{1},p_{2})
    \mathcal{U}_{\hbox{\tiny{sym}}}(\theta^{\prime},\phi^{\prime},\psi^{\prime},q_{1}^{\prime},q_{2}^{\prime},p_{1}^{\prime},p_{2}^{\prime})f)(r_{1},r_{2}),
    \end{align}
    with $f\in L^{2}(\mathbb{R}^{2}, dr_{1}dr_{2})$.

The adjoint of $\mathcal{U}_{\hbox{\tiny{sym}}}$ now reads
    \begin{eqnarray}\label{adjoint-symmetric-gauge-rep}
    \lefteqn{(\mathcal{U}^{*}_{\hbox{\tiny{sym}}}(\theta,\phi,\psi,q_{1},q_{2},p_{1},p_{2})f)(r_{1},r_{2})}\nonumber\\
    &&=e^{-i(\theta+\phi+\psi)}e^{-i\left[\alpha p_{1}r_{1}+\alpha p_{2}r_{2}-\frac{\alpha(\alpha-\sqrt{\alpha^{2}-\beta\gamma})}{\beta}(q_{1}r_{2}-q_{2}r_{1})+\frac{\sqrt{\alpha^{2}-\beta\gamma}}{2}(p_{1}q_{1}+p_{2}q_{2})\right]}\nonumber\\
    &&\times f\left(r_{1}+\frac{\beta}{2\alpha}p_{2}-\frac{\alpha+\sqrt{\alpha^{2}-\beta\gamma}}{2\alpha}q_{1},r_{2}-\frac{\beta}{2\alpha}p_{1}-\frac{\alpha+\sqrt{\alpha^{2}-\beta\gamma}}{2\alpha}q_{2}\right),\label{adjoint-sym-rep-exprssn}
    \end{eqnarray}
    from which unitarity of $\mathcal{U}_{\hbox{\tiny{sym}}}$ follows immediately
    \begin{equation}\label{unitarity-condtn}
    (\mathcal{U}_{\hbox{\tiny{sym}}}\mathcal{U}^{*}_{\hbox{\tiny{sym}}}f)(r_{1},r_{2})=(\mathcal{U}^{*}_{\hbox{\tiny{sym}}}\mathcal{U}{\hbox{\tiny{sym}}}f)(r_{1},r_{2})=f(r_{1},r_{2}),
    \end{equation}
    where $f\in L^{2}(\mathbb{R}^{2},dr_{1}dr_{2})$.

    It remains to prove the irreducibility of the unitary representation $\mathcal{U}_{\hbox{\tiny{sym}}}$ of the Lie group $\g$. Note that the representation of the corresponding Lie algebra $\G$ given by (\ref{sym-gauge-algbr-rep}) is irreducible (see the discussion on its irreducibility in section (\ref{sec:gauges-ncqm})). Also, $\g$ is a connected, simply connected Lie group, the corresponding representation of the Lie group is also irreducible. But the equivalence classes of unitary irreducible representations of $\g$ are all obtained in Section \ref{sec:coad-orbts}. And the group representation complying with the commutation relations (\ref{commutators-rep-equivalent-nonzero}) is given by (\ref{nonzero-rep}). Therefore, the unitary irreducible representation of $\g$, due to the choice of symmetric gauge of vector potential, has to be equivalent to one of the representations (\ref{nonzero-rep}) computed in the Hilbert space $L^{2}(\mathbb{R}^{2},dr_{1}dr_{2})$, for some nonzero value of $\rho$, $\sigma$, and $\tau$. Comparing (\ref{symmetric-gauge-rep}) with (\ref{nonzero-equivalent-rep}), one finds the underlying coadjoint orbit coordinates correspond to $\rho=\sigma=\tau=1$. Therefore, the unitary representation (\ref{symmetric-gauge-rep}) of $\g$ is indeed irreducible.
\qed

\bigskip
\prf {\bf of Theorem \ref{thm-trnsfrm-matrix}}

We first write $\mathbb{M}$ in the block-form:
\begin{equation}\label{blk-frm-exprsn}
   \mathbb{M}=\begin{bmatrix}A&B\\C&D\end{bmatrix}=\begin{bmatrix}A_{11}&A_{12}&B_{11}&B_{12}\\A_{21}&A_{22}&B_{21}&B_{22}\\C_{11}&C_{12}&D_{11}&D_{12}\\C_{21}&C_{22}&D_{21}&D_{22}\end{bmatrix}
    \end{equation}
    (\ref{trnsfrmtn-matrx-def}) then yields
    \begin{eqnarray}\label{first-eq}
    \begin{bmatrix}A_{11}&A_{12}\\A_{21}&A_{22}\end{bmatrix}\begin{bmatrix}\hat{Q}_{1}\\\hat{P}_{2}\end{bmatrix}+\begin{bmatrix}B_{11}&B_{12}\\B_{21}&B_{22}\end{bmatrix}\begin{bmatrix}\hat{Q}_{2}\\\hat{P}_{1}\end{bmatrix}&=&\begin{bmatrix}\hat{Q}^{\prime}_{1}\\\hat{P}^{\prime}_{2}\end{bmatrix}\nonumber\\
    \implies\begin{bmatrix}A_{11}\hat{Q}_{1}+A_{12}\hat{P}_{2}+B_{11}\hat{Q}_{2}+B_{12}\hat{P}_{1}\\A_{21}\hat{Q}_{1}+A_{22}\hat{P}_{2}+B_{21}\hat{Q}_{2}+B_{22}\hat{P}_{1}\end{bmatrix}&=&\begin{bmatrix}\hat{Q}^{\prime}_{1}\\\hat{P}^{\prime}_{2}\end{bmatrix}.
    \end{eqnarray}
    Similarly
    \begin{eqnarray}\label{second-eq}
    \begin{bmatrix}C_{11}&C_{12}\\C_{21}&C_{22}\end{bmatrix}\begin{bmatrix}\hat{Q}_{1}\\\hat{P}_{2}\end{bmatrix}+\begin{bmatrix}D_{11}&D_{12}\\D_{21}&D_{22}\end{bmatrix}\begin{bmatrix}\hat{Q}_{2}\\\hat{P}_{1}\end{bmatrix}&=&\begin{bmatrix}\hat{Q}^{\prime}_{2}\\\hat{P}^{\prime}_{1}\end{bmatrix}\nonumber\\
    \implies\begin{bmatrix}C_{11}\hat{Q}_{1}+C_{12}\hat{P}_{2}+D_{11}\hat{Q}_{2}+D_{12}\hat{P}_{1}\\C_{21}\hat{Q}_{1}+C_{22}\hat{P}_{2}+D_{21}\hat{Q}_{2}+D_{22}\hat{P}_{1}\end{bmatrix}&=&\begin{bmatrix}\hat{Q}^{\prime}_{2}\\\hat{P}^{\prime}_{1}\end{bmatrix}.
    \end{eqnarray}
    Using (\ref{first-eq}) and (\ref{second-eq}), one gets
    \begin{eqnarray}\label{commutn-first}
    \lefteqn{[\hat{Q}^{\prime}_{1},\hat{P}^{\prime}_{1}]}\nonumber\\
    &&=[A_{11}\hat{Q}_{1}+A_{12}\hat{P}_{2}+B_{11}\hat{Q}_{2}+B_{12}\hat{P}_{1},C_{21}\hat{Q}_{1}+C_{22}\hat{P}_{2}+D_{21}\hat{Q}_{2}+D_{22}\hat{P}_{1}]\nonumber\\
    &&=A_{11}D_{21}(i\vartheta\mathbb{I})+A_{11}D_{22}(i\hbar\mathbb{I})+A_{12}D_{21}(-i\hbar\mathbb{I})+A_{12}D_{22}(-i\bm{\mathcal{B}}\mathbb{I})\nonumber\\
    &&+B_{11}C_{21}(-i\vartheta\mathbb{I})+B_{11}C_{22}(i\hbar\mathbb{I})+B_{21}C_{21}(-i\hbar\mathbb{I})+B_{12}C_{22}(i\bm{\mathcal{B}}\mathbb{I})\nonumber\\
    &&=i\hbar(A_{11}D_{22}-A_{12}D_{21}+B_{11}C_{22}-B_{12}C_{21})\mathbb{I}+i\vartheta(A_{11}D_{21}-B_{11}C_{21})\mathbb{I}\nonumber\\
    &&+i\bm{\mathcal{B}}(B_{12}C_{22}-A_{12}D_{22})\mathbb{I}.
    \end{eqnarray}
    But we are given that $[\hat{Q}^{\prime}_{1},\hat{P}^{\prime}_{1}]=i\hbar\mathbb{I}$. Therefore, (\ref{commutn-first}) reduces to
    \begin{equation}\label{com-first-final-exprssn}
    \frac{\vartheta}{\hbar}(B_{11}C_{21}-A_{11}D_{21})+\frac{\bm{\mathcal{B}}}{\hbar}(A_{12}D_{22}-B_{12}C_{22})+(A_{12}D_{21}+B_{12}C_{21}-A_{11}D_{22}-B_{11}C_{22})=-1.
    \end{equation}
    In exactly the same way one can go on to compute $[\hat{Q}^{\prime}_{2},\hat{P}^{\prime}_{2}]$, $[\hat{Q}^{\prime}_{1},\hat{Q}^{\prime}_{2}]$, $[\hat{P}^{\prime}_{1},\hat{P}^{\prime}_{2}]$, $[\hat{Q}^{\prime}_{1},\hat{P}^{\prime}_{2}]$, and $[\hat{Q}^{\prime}_{2},\hat{P}^{\prime}_{1}]$ and thereby obtain the following set of equations:
    \begin{equation}\label{other-com-final-exprssn}
    \begin{split}
    &\frac{\vartheta}{\hbar}(B_{21}C_{11}-A_{21}D_{11})+\frac{\bm{\mathcal{B}}}{\hbar}(A_{22}D_{12}-B_{22}C_{12})\\
    &\qquad+(B_{22}C_{11}+A_{22}D_{11}-B_{21}C_{12}-A_{21}D_{12})=1,\\
    &(A_{12}D_{11}+B_{12}C_{11}-A_{11}D_{12}-B_{11}C_{12})+\frac{\vartheta}{\hbar}(B_{11}C_{11}-A_{11}D_{11})\\
    &\qquad+\frac{\bm{\mathcal{B}}}{\hbar}(A_{12}D_{12}-B_{12}C_{12})=-\frac{\vartheta}{\hbar},\\
    &(B_{22}C_{21}+A_{22}D_{21}-B_{21}C_{22}-A_{21}D_{22})+\frac{\bm{\mathcal{B}}}{\hbar}(A_{22}D_{22}-B_{22}C_{22})\\
    &\qquad+\frac{\vartheta}{\hbar}(B_{21}C_{21}-A_{21}D_{21})=\frac{\bm{\mathcal{B}}}{\hbar},\\
    &\frac{\vartheta}{\hbar}(A_{11}B_{21}-A_{21}B_{11})+\frac{\bm{\mathcal{B}}}{\hbar}(A_{22}B_{12}-A_{12}B_{22})\\
    &\qquad+(A_{11}B_{22}+A_{22}B_{11}-A_{12}B_{21}-A_{21}B_{12})=0,\\
    &\frac{\vartheta}{\hbar}(C_{11}D_{21}-C_{21}D_{11})+\frac{\bm{\mathcal{B}}}{\hbar}(D_{12}C_{22}-C_{12}D_{22})\\
    &\qquad+(C_{11}D_{22}-C_{12}D_{21}+D_{11}C_{22}-D_{12}C_{21})=0.
    \end{split}
    \end{equation}
    Now (\ref{com-first-final-exprssn}) and the set of relations enumerated in (\ref{other-com-final-exprssn}) can all be compactified into the following three matrix equations:
    \begin{equation}\label{compctified-matrx-eqn}
    \begin{split}
    &AQB^{T}-BQ^{T}A^{T}=0,\\
    &CQD^{T}-DQ^{T}C^{T}=0,\\
    &AQD^{T}-BQ^{T}C^{T}=Q,
    \end{split}
    \end{equation}
    where $Q$ is the $2\times 2$ matrix given by (\ref{two-by-two-matrx}).

    The three matrix equations (\ref{compctified-matrx-eqn}) can yet be incorporated in one single $4\times 4$ matrix equation given by
    \begin{equation*}
    \begin{bmatrix}A&B\\C&D\end{bmatrix}\begin{bmatrix}0&Q\\-Q^{T}&0\end{bmatrix}\begin{bmatrix}A^{T}&C^{T}\\B^{T}&D^{T}\end{bmatrix}=\begin{bmatrix}0&Q\\-Q^{T}&0\end{bmatrix},
    \end{equation*}
    which boils down to
    \begin{equation*}
    \mathbb{M}\mathbb{Q}\mathbb{M}^{T}=\mathbb{Q}.
    \end{equation*}
    \qed
\section*{Acknowledgements} 
The authors would like to thank the referees for numerous insightful comments which helped to clarify several points in the paper and to close some gaps, thereby considerably improving the presentation. One of them (STA) acknowledges a grant from the Natural Science and Engineering Research Council of Canada (NSERC).


\begin{thebibliography}{99}

\bibitem{acat}
	C. Acatrinei,
	\newblock {\em Path integral formulation of noncommutative quantum mechanics. }
 \newblock {J. High Energy Phys.}, {\bf 09}, (2001) 007. doi:10.1088/1126-6708/2001/09/007.

\bibitem{aliismailshah}
S. T. Ali, M. E. H. Ismail and Nurisya M. Shah,
\newblock {\em Deformed complex Hermite polynomials,}
\newblock {to appear.}

\bibitem{alibaloghshah}
F. Balogh, S.T. Ali  and Nurisya M. Shah,
\newblock {\em Some biorthogonal families of polynomials arising in noncommutative quantum mechanics,}
\newblock {arXiv:1309.4163v1 [math-ph] 17 Sep 2013}

\bibitem{bas}
	C. Bastos, O. Bertolami, N.C. Dias and J.N. Prata,
	\newblock {\em Weyl-Wigner formulation of noncommutative quantum mechanics.}
 \newblock{J. Math. Phys.}, {\bf 49}, (2008) 072101.


\bibitem{ncqmjmp}
S. H. H. Chowdhury and S. T. Ali,
\newblock {\em The symmetry groups of noncommutative quantum mechanics and coherent state quantization.}
\newblock {\em J. Math. Phys.}, {\bf 54}, 032101 (2013).


\bibitem{delducetalt}
F. Delduc, Q. Duret, F. Gieres and M. Lafran\c{c}ois,
\newblock {\em Magnetic fields in noncommutative quantum mechanics\/,}
\newblock {J. Phys. Conf. Ser.} {\bf 103}, 012020 (2008).

\bibitem{dias}
	N.C. Dias, M. Gosson, F. Luef and J.N. Prata,
	\newblock {\em A deformation quantization theory for noncommutative quantum mechanics.}
\newblock {J. Math. Phys.},  {\bf 51}, (2010) 072101. doi:10.1063/1.3436581.

\bibitem{gamboa1}
	J. Gamboa, M. Loewe  and J.C. Rojas,
	\newblock {\em Non-commutative Quantum Mechanics.}
   \newblock{ Phys. Rev. D}, {\bf 64}, (2001) 067901. doi: 10.1103/PhysRevD.64.067901

\bibitem{ghanmi}
Ghanmi, A.,
\newblock {\em A class of generalized complex Hermite polynomials.}
\newblock {J. Math. Anal. and App.} {\bf 340} (2008), 1395-1406.

\bibitem{hormarsti} Horv\'athy, P. A., Martina, L., and Stichel, P. C.,
  \newblock {\em Exotic Galilean symmetry and non-commutative mechanics.}
  \newblock {Symmetry, Integr. Geom.: Methods Appl.},  {\bf 6}, (2010)  060.

\bibitem{intissar}
A. Intissar, and A. Intissar,
\newblock {\em Spectral properties of the Cauchy transform on $L_{2}(\mathbb{C}, e^{-\left| z\right|^{2}}\lambda(z))$,}
\newblock {J. Math. Anal. and Applications.} {\bf 313} (2006), 400--418.

\bibitem{Ito} K. Ito,
\newblock {\em Complex multiple Wiener integral,}
\newblock {Japan J. Math.} {\bf 22}(1952), 63--86.

\bibitem{Kirillovbook}
A. A. Kirillov,
\newblock {\em Lectures on the Orbit Method},
\newblock American Math. Soc., 2004.


\bibitem{scho2}
	F.G. Scholtz, B. Chakraborty, S. Gangopadhyay and A.G. Hazra,
	\newblock {Dual families of non-commutative quantum system.}
 \newblock {Phys. Rev. D}, {\bf 71}, (2005) 085005.
	

\bibitem{scho}
	F.G. Scholtz, L. Gouba, A. Hafver and C.M. Rohwer,
	\newblock {\em Formulation, interpretation and application of non-commutative quantum mechanics.}
\newblock {J. Phys. A},  {\bf 42},  (2009) 175303. doi:10.1088/1751-8113/42/17/175303 175303.

\bibitem{Shigekawa}
I. Shigekawa,
\newblock {\em Eigenvalue problems for the Schr\"odinger operators with magnetic field on a compact Riemannian manifold.}
\newblock {J. Funct. Anal.}, {\bf 75}, no. 1 (1987) 92--127.


\end{thebibliography}
    \end{document}